\documentclass[prd,twocolumn,showpacs,amsmath,amssymb,floatfix,superscriptaddress,nofootinbib]{revtex4}
\usepackage[colorlinks=true, citecolor=blue, linkcolor=blue, urlcolor=blue]{hyperref}
\usepackage{amsfonts,bm,color,dcolumn,xcolor,graphicx,amsmath,multirow}
\usepackage{amssymb,latexsym,subfigure,threeparttable,multirow,makecell,tabularx,setspace,textcomp}

\allowdisplaybreaks

\newcommand{\DS}[1]{/\!\!\!#1}

\begin{document}

\title{$a_0(980)$-meson twist-2 distribution amplitude within the QCD sum rules and investigation of $D \to a_0(980) (\to\eta\pi) e^+\nu_e$}

\author{Zai-Hui Wu}
\author{Hai-Bing Fu\footnote{Corresponding author}}
\email{fuhb@cqu.edu.cn}
\author{Tao Zhong}
\email{zhongtao1219@sina.com}
\author{Dong Huang}
\address{Department of Physics, Guizhou Minzu University, Guiyang 550025, China}
\author{Dan-Dan Hu}
\author{Xing-Gang Wu}
\email{wuxg@cqu.edu.cn}
\address{Department of Physics, Chongqing Key Laboratory for Strongly Coupled Physics, Chongqing University, Chongqing 401331, China}

\date{\today}

\begin{abstract}
\begin{spacing}{1.1}
In this paper, moments of $a_0(980)$-meson twist-2 light-cone distribution amplitudes were deeply researched by using QCD sum rules approach within background field theory. Up to 9th-order accuracy, we present $\langle\xi_{2;a_0}^n\rangle|_{\mu_0}$ at the initial scale $\mu_0 = 1~{\rm GeV}$, i.e. $\langle\xi^1_{2;a_0}\rangle|_{\mu_0} = -0.307(43)$, $\langle\xi^3_{2;a_0}\rangle|_{\mu_0} = -0.181(34)$, $\langle\xi^5_{2;a_0}\rangle|_{\mu_0} = -0.078(28)$, $\langle\xi^7_{2;a_0}\rangle|_{\mu_0} = -0.049(26)$, $\langle\xi^9_{2;a_0}\rangle|_{\mu_0} = -0.036(24)$, respectively. An improved light-cone harmonic oscillator model for $a_0(980)$-meson twist-2 light-cone distribution amplitudes is adopted, where its parameters are fixed by using the least squares method based on the $\langle\xi_{2;a_0}^n\rangle|_{\mu_0}$, and their goodness of fit reach to $95.4\%$. Then, we calculate the $D\to a_0(980)$ transition form factors within the light-cone sum rules approach, and at largest recoil point, we obtain $f_+^{D\to a_0}(0) = 1.058^{+0.068}_{-0.035}$ and $f_-^{D\to a_0}(0) = 0.764^{+0.044}_{-0.036}$. As a further application, the branching fractions of the $D\to a_0(980)\ell\bar\nu_\ell$ semileptonic decays are given. Taking the decay $a_0(980)\to \eta\pi$ into consideration, we obtain ${\cal B}(D^0 \to a_0(980)^- (\to \eta \pi^-) e^+\nu_e) =(1.330^{+0.216}_{-0.134})\times10^{-4}$, ${\cal B}(D^+\to a_0(980)^0(\to \eta \pi^0)e^+\nu_e)=(1.675^{+0.272}_{-0.169})\times10^{-4}$, which are consistent with the BESIII collaboration and PDG data within errors. Finally, we present the angle observables of forward-backward asymmetries, $q^2$-differential flat terms and lepton polarization asymmetry of the semileptonic decay $D\to a_0(980)\ell\bar\nu_\ell$.
\end{spacing}
\end{abstract}

\maketitle

\section{Introduction}

In the past few decades, especially for the discovery of resonance $a_0(980)$ state~\cite{Ammar:1968zur}, nature of scalar mesons below 1 GeV is a long-standing puzzle, which becomes one of hot topics in hadron physics. In the quark model scenario, the composition of $a_0(980)$ has turned out to be mysterious. Its intriguing internal structure allows tests of various hypotheses, such as quark-antiquark~\cite{Achasov:1987ts,Cheng:2005nb}, tetraquark states~\cite{Jaffe:1976ig, Alford:2000mm, Humanic:2022hpq, Brito:2004tv, Klempt:2007cp, Alexandrou:2017itd}, two-meson molecule bound states~\cite{Weinstein:1982gc, Branz:2007xp, Dai:2012kf, Dai:2014lza, Sekihara:2014qxa} and hybrid states~\cite{Ishida:1995}. However, until now there is no definite conclusion of which scenario is correct or have no general agreement on the inner structure of $a_0(980)$. Among the hadronic decay processes including $a_0(980)$ state, semileptonic decay of charmed meson provides a simpler decay mechanism and final-state interactions, which can give an ideal platform for studying the meson's properties.

From the experimental side, BESIII collaboration observed the charmless hadronic decay processes involving $a_0(980)$-meson, i.e. $D^0\to a_0 (980)^- e^+\nu_e$ and $D^+\to a_0 (980)^0e^+ \nu_e$ with the significance up to $6.4\sigma$ and $2.9\sigma$, respectively~\cite{BESIII:2018sjg}. Experimental facilities have reported the most precise results on the semileptonic decay of $D_{(s)}$ to pseudoscalar and vector mesons. From theory point of view, these channels are straightforward to study because the internal stucture/quark content of the meson is a typical quark-antiquark system. But the quark structure of the scalar meson below 1 GeV has varied explanations, cf. the review ``Scalar mesons below 1 GeV'' of Particle Data Group (PDG)~\cite{ParticleDataGroup:2022pth}. Wang {\it et al.} conclude the ratio $(R)$ provides a model independent way to distinguish the quark components of the light scalar meson, i.e. the four-quark and two-quark pictures~\cite{Wang:2009azc}. And this value has not confirmed exactly from experiments. Nowadays, the two-quark picture are researched by some theoretical group, such as covariant confined quark model (CCQM)~\cite{Soni:2020sgn}, light-cone sum rule (LCSR)~\cite{Cheng:2017fkw,Huang:2021owr}, AdS/QCD~\cite{Momeni:2022gqb},  perturbative QCD (pQCD)~\cite{Rui:2018mxc} and Bethe-Salpeter equation~\cite{Santowsky:2021ugd}. So it is also meaningful to reconsider the two-quark scenario in detail and to compare the experimental observable, which is also the starting point of this paper.

The important physical quantity for semileptonic decays $D\to a_0(980)\ell\bar \nu_\ell$ is the transition form factors (TFFs). As is known that the analogous semileptonic TFFs are also the essential ingredients for the indirect search of new physics beyond the Standard Model~\cite{Aslam:2009cv}. Therefore, an accurate TFFs is crucial for the semileptonic decay process. The LCSR approach is one of an effective method in dealing with heavy to light decays, which makes the operator production expansion (OPE) into the increasing twist light-cone distribution amplitudes (LCDAs), i.e twist-2, 3, 4 LCDAs. The LCSR is suitable in the large and middle recoil region. In this paper, we will adopt the LCSR to recalculate the $D\to a_0(980)$ TFFs and the experimental observable. One of the key quantities that characterize the TFFs is the twist-2 LCDA, which describes the dominant momentum fraction distribution for each part of a meson. In general, the $a_0(980)$-meson twist-2 LCDA $\phi_{2;a_0}(x,\mu)$ can be expanded as a Gegenbauer polynomial series
\begin{eqnarray}
\phi_{2;a_0}(x,\mu) = 6x\bar x \bigg[a_{2;a_0}^0(\mu)  + \sum\limits_{n=1}^\infty a_{2;a_0}^n(\mu) C_n^{3/2}(\xi) \bigg],
\end{eqnarray}
where $\bar x = (1-x)$ and $\xi = (2x-1)$. The zeroth-order Gegenbauer moment $a_{2;a_0}^0(\mu)$ is equals to zero, and the even Gegenbauer coefficients are highly suppressed, which exactly equal to zero under the approximation that $m_1\simeq m_2$ ($m_{1,2}$ are masses of two constituent quarks), and the LCDA of the scalar meson is then dominated by the odd Gegenabuer moments. In contrast, the odd Gegenbauer moments vanish for $\pi$ and $\rho$ mesons. Thus, the behavior of $\phi_{2;a_0}(x,\mu)$ tends to antisymmetric form under the exchange $u\to (1-u)$ in isospin symmetry. Currently, the $a_0(980)$-meson twist-2 LCDA is mainly coming from QCDSR by Cheng {\it et al.}~\cite{Cheng:2005nb}, which gives the first two nonzero $\xi$-moments~\footnote{The $\xi$-moments can be related with Gegenbauer moments directly, which the formulae can be found in Ref.~\cite{Cheng:2005nb}}. In our previous work~\cite{Zhong:2022lmn}, based on the pionic leading-twist DA, we analyzed in detail the influence of different numbers of $\xi$-moments included in the fitting, and found that when the order of $\xi$-moments is not more than ten, the change of the number of $\xi$-moments has an obvious impact on the fitting result. When the order of $\xi$-moment is more than ten, the change of the number of $\xi$-moments has a very small impact on the fitting results. Thus, the higher-order $\xi$-moments should be given in order to get more accuracy $\phi_{2;a_0}(x,\mu)$ behavior and get more accuracy predictions for the processes involving $a_0(980)$-meson.

To achieve this target, the QCDSR under the framework of background field theory (BFT) is
one of the effective way~\cite{Novikov:1983gd, Hubschmid:1982pa, Govaerts:1983ka, Govaerts:1984bk, Ambjorn:1982bp, Ambjorn:1982en, Reinders:1984sr, Elias:1987ac, Huang:1986wm, Huang:1989gv}. In this approach, the quark and gluon fields are composed by background fields and quantum fluctuations around them. By decomposing quark and gluon fields into classical background fields describing nonperturbative effects and quantum fields describing perturbative effects, BFT can provide clear physical images for the separation of long-range and short-range dynamics of OPE. At present, the BFT has been applied to calculate the LCDAs of pseudoscalar and vector/axial vector mesons~\cite{Fu:2016yzx,Hu:2021zmy, Hu:2021lkl, Zhong:2014jla, Fu:2018vap, Zhong:2016kuv, Huang:2004tp, Zhong:2011rg}. Therefore, we will study the scalar $a_0(980)$-meson twist-2 LCDA by using BFT. To get a more accurate behavior of $\phi_{2;a_0}(x,\mu)$, a research scheme is used which combine the LCHO model and nonperturbative QCD sum rule for $\xi$-moments. Specifically, a new QCD sum rule formula will be used due to the zeroth-order $\xi$-moments can not be normalized in the whole Borel parameters region. In this paper, we will calculate the first five-order nonzero $\xi$-moments and determine the LCHO model parameters with the least squares method. This scheme is used in the pion case~\cite{Zhong:2021epq}, and subsequently for the kaon leading-twist DA~\cite{Zhong:2022ecl} and $a_1(1260)$-meson longitudinal twist-2 DA~\cite{Hu:2021lkl}.

The remaining parts of this paper are organized as follows. In Sec.~\ref{Sec:II}, we calculate the $a_0(980)$-meson twist-2 LCDA moments and introduce the LCHO model. A new improved model is proposed and the model parameters shall be obtained by fitting moment with the least square method. Section~\ref{Sec:III} gives the numerical results and discussions, which include the transition form factor, decay width, decay branching ratio of semileptonic decays $D\to a_0(980)\ell\bar \nu_\ell$. Meanwhile, the forward-backward asymmetries, the $q^2$ differential flat terms and lepton polarization asymmetry of the semileptonic decay $D\to a_0(980)\ell\bar\nu_\ell$ are also given. Section~\ref{Sec:IV} is for a brief summary.

\section{Theoretical framework}\label{Sec:II}

The $a_0(980)$-meson have three types of states, one is $a_0(980)^0$-meson with $(\bar uu-\bar dd)/\sqrt 2$ component, the other is $a_0(980)^-$-meson with $(\bar ud)$ component, and the third one is $a_0(980)^+$-meson with $(\bar du)$. Based on the basic procedure of QCD sum rules, one can adopt the following two-point correlator to derive the sum rules for $a_0(980)$-meson twist-2 LCDA moments $\langle\xi^n_{2;a_0}\rangle|_\mu$, which can be read off,
\begin{eqnarray}
\Pi^{(n,0)}_{2;a_0} (z,q) &=& i \int d^4x e^{iq\cdot x} \langle 0| T \{ J_n^V(x), J^{S,\dag}_0(0) \} |0\rangle
\nonumber\\
&=& (z\cdot q)^{n+1} I_{2;a_0} (q^2),
\label{Eq:correlator}
\end{eqnarray}
where $z^2 = 0$ and $n$ take odd numbers. The currents are taken as $J^V_n(x) = \bar q_1(x) \DS z (i z\cdot \tensor{D})^n q_2(x)$ and $J^S_0(0) = \bar q_1(0)q_2(0)$, which are mainly come from definitions of the scalar $a_0(980)$-meson twist-2 LCDA $\phi_{2;a_0}(x,\mu)$ and twist-3 LCDA $\phi_{3,a_0}^p (x,\mu)$, which can be written as~\cite{Cheng:2005nb}
\begin{align}
&\hspace{-0.2cm}\langle 0|\bar q_1(z)\gamma _\mu q_2(-z)|a_0(p)\rangle
\nonumber\\
&\qquad\qquad = p_\mu \bar f_{a_0} \int^1_0 d xe^{i(2u-1)(p\cdot z)}\phi_{2;a_0}(x,\mu),
\label{Eq:a0}
\\
&\hspace{-0.2cm}\langle 0|\bar q_1(z) q_2(-z)|a_0(p)\rangle
\nonumber\\
&\qquad\qquad = m_{a_0}\bar f_{a_0} \int^1_0 dx e^{i(2u-1)(p\cdot z)}\phi_{3,a_0}^p(x,\mu),
\label{Eq:phi3s}
\end{align}
From one side, one can apply the OPE for the correlator, e.g. Eq.~\eqref{Eq:correlator} in deep Euclidean region $q^2 \ll 0$.  The calculation is carried out in the framework of BFT~\cite{Huang:1989gv}. Based on the basic assumptions and Feynman rules of BFT, the correlator can be expanded into three terms including the quark propagators and the vertex operators,

\begin{align}
\Pi^{(n,0)}_{2;a_0}(z,q)&= i\int d^4x e^{iq\cdot x}\Big\{
\nonumber\\
&-{\rm Tr}\langle 0|S^{q_1}_{F}(0,x)\DS z (iz\cdot\tensor D)^n S^{q_2}_F(x,0)|0\rangle
\nonumber\\
& +{\rm Tr}\langle 0|\bar q_1(x)q_1(0)\DS z(iz\cdot\tensor{D})^n S^{q_2}_F(x,0)|0\rangle
\nonumber\\
& +{\rm Tr}\langle 0|S^{q_1}_F(0,x)\DS z(iz\cdot\tensor{D})^n \bar q_2(x)q_2(0)|0\rangle
\nonumber\\
& +\cdots \Big\}
\label{pp}
\end{align}
where ``Tr'' indicates trace for the $\gamma$-matrix and color matrix, $S^{q_1}_{F}(0,x)$ and $S^{q_2}_{F}(x,0)$ indicate the light-quark propagator from $x$ to 0 and from 0 to $x$.  The $\DS z (iz\cdot\tensor D)^n$ are the vertex operators from current $J_n(x)$. The expressions up to dimension-six of the quark propagator, the vertex operator, and the vacuum matrix elements such as $\langle 0|\bar q_1(x)q_1(0)|0\rangle$ and $\langle 0|\bar q_2(x)q_2(0)|0\rangle$ have been derived in our previous works~\cite{Zhong:2011rg, Zhong:2014jla, Zhong:2021epq, Hu:2021zmy}. By substituting those corresponding formula into Eq.~\eqref{pp}, the OPE of correlator~\eqref{Eq:correlator}, $I^{\rm QCD}_{2;a_0} (q^2)$, can be obtained. Meanwhlie, we take $q_1=q_2=q$ stand for the light $u$, $d$-quark. On the other hand, one can insert a complete set of $a_0(980)$-meson intermediated hadronic states with the same $J^P$ quantum number into the correlator and obtain the hadronic expression
\begin{align}
\textrm{Im} I^{(n,0)}_{2;a_0,{\rm had}}(q^2)&=\pi \delta (q^2 - m^{2}_{a_0}) m_{a_0} \bar f^{2}_{a_0} \langle\xi^n_{2;a_0}\rangle|_\mu\langle \xi^{p;0}_{3;a_0}\rangle|_\mu
\nonumber\\
& - \frac{3m_q}{4\pi^2(n+2)}\theta(q^2- s_{a_0}).
\label{hadim}
\end{align}
with $m_q = m_{q_1} = m_{q_2}$ is the light quark mass, where the slight difference between mass of $u$-quark and $d$-quark is ignored in this paper. Meanwhile, the $s_{a_0}$ stands for the continuum threshold. Here the definitions of moments for $a_0(980)$-meson twist-2, 3 LCDAs are used, which have the following formula
\begin{align}
& \hspace{-0.3cm}\langle 0 |\bar q_1 (0) \DS z (iz \cdot\tensor D)^n q_2(0)|a_0(p)\rangle = (z\cdot p)^{n+1}\bar f_{a_0}\langle\xi^n_{2;a_0}\rangle|_\mu,
\label{DA2}\\
& \hspace{-0.3cm} \langle 0|\bar q_1(0) q_2(0)|a_0(p)\rangle = m_{a_0} \bar f_{a_0} \langle \xi^{p,0}_{3;a_0} \rangle|_\mu.
\label{DA3}
\end{align}
where $m_{a_0}$ and $\bar f_{a_0}$ stand for $a_0(980)$-meson mass and its decay constant, respectively. By considering the dispersion relation and performing Borel transformation, we can get the sum rule for $\langle\xi_{2;a_0}^n\rangle|_\mu \langle\xi_{3;a_0}^{p;0}\rangle|_\mu$,
\begin{widetext}
\begin{eqnarray}
&&\hspace{-0.3cm}\frac{\langle\xi^n_{2;a_0}\rangle|_\mu\langle \xi^{p;0}_{3;a_0}\rangle|_\mu m_{a_0} \bar f_{a_0}^2} { e^{m_{a_0}^2/M^2}} = - \frac{3 m_q}{4\pi^2(n+2)} \left(1-e^{-s_0/M^2}\right) + 2 \langle\bar qq \rangle + \frac{\langle g_s\bar q q \rangle^2 }{81M^4} 4m_q (n+3)  -  \frac{\langle g_s^2\bar qq\rangle^2}{1944 \pi^2 M^4} m_q(2+\kappa^2)
\nonumber\\
&& \qquad \times \bigg\{\delta^{0n}\Big[-24\,\bigg(-\ln\frac{M^2}{\mu^2}\bigg)-148\Big]\,+\,\delta^{1n}\bigg[128 \bigg(-\ln\frac{M^2}{\mu^2}\bigg)- 692\bigg] \,+\, \theta(n-1)\bigg[8\,(6n^2+34n)\,\bigg(- \ln\frac{M^2}{\mu^2}\bigg)
\nonumber\\
&& \qquad+4n\tilde\psi(n)-2(6n^2+96n+212)\bigg]+\theta(n-2)\bigg[8(33n^2-17n)
\bigg(-\ln\frac{M^2}{\mu^2}\bigg) - 2(6n^2 +71n)\tilde\psi(n) - \frac{1}{n(n-1)}
\nonumber\\
&& \qquad \times  (231n^4\,+520n^3\,-1101n^2\,+230n)\,\bigg]\, +\, \theta(n-3) \bigg[(74n-144n^2)\, \tilde\psi(n) \,-\, \frac{1}{n-1}\, (169n^3-348n^2+245n
\nonumber\\
&& \qquad +60)\bigg]+4(n+5)\bigg\} \,-  \frac{\langle\alpha_s G^2\rangle}{24\pi M^2} m_q \bigg\{ 12n \bigg( -\ln\frac{M^2}{\mu^2} \bigg) -6(n+2) +\theta(n-1) \bigg[4n\bigg(-\ln\frac{M^2}{\mu^2} \bigg) + 3\tilde\psi(n)-\frac6n\bigg]
\nonumber\\
&&\qquad
+\theta(n-2)\bigg[-(8n+3) \tilde \psi(n) - 2(2n+1) + \frac6n \bigg] \bigg\} -  \frac{ \langle g_s^3 f G^3 \rangle}{192 \pi^2 M^4} m_q  \bigg\{ \delta^{1n} \bigg[-24 \bigg( - \ln \frac{M^2} {\mu^2} \bigg)+84\bigg]+\theta(n-1)
\nonumber\\
&&\qquad
\times\bigg[-4n(3n-5)\,\bigg(-\ln\frac{M^2}{\mu^2}\bigg) \,+\, 2 (2n^2 \,+5n \, -13)\bigg]\,+\,\theta(n-2)\,\bigg[-24n^2\bigg(-\ln\frac{M^2}{\mu^2}\bigg) \,+\, 2n\,(n-4) \, \tilde\psi(n)
\nonumber\\
&&\qquad  + 17n^2 +55n+12\bigg] +\theta(n-3)\bigg[2n(n-4)\tilde\psi(n)+\frac{1}{n-1}(19n^3-32n^2+7n+6)\bigg] \bigg\} -  \frac{\langle{g_s}\bar q \sigma TGq\rangle }{3M^2} 4n,
\label{Eq:SRxi2nxi30}
\end{eqnarray}
\end{widetext}
with $\tilde\psi(n)=\psi(\frac{n+1}2) - \psi(\frac n2) +(-1)^n\ln4$. For a more thorough considering the sum rule for $\langle\xi^n_{2;a_0}\rangle|_\mu$, it might be convenient to calculate $\langle \xi^{p;0}_{3;a_0}\rangle|_\mu$. To achieve this target, one can use the correlator $\Pi^{(0,0)}_{3;a_0} (z,q) = i \int d^4x e^{iq\cdot x} \langle 0| T \{ J^S_0(x), J^{S\dag}_0(0)\} |0\rangle$ for the $a_0(980)$-meson twist-3 LCDA 0th moment. Followed by the basic procedure of the QCDSR, the expression of the 0th moment of scalar meson $a_0(980)$-meson twist-3 LCDA can be obtained,
\begin{align}
    &\frac{(\langle \xi^{p;0}_{3;a_0}\rangle|_\mu)^2 m_{a_0}^2 \bar f_{a_0}^2}{M^2e^{m_{a_0}^2/M^2}}\!=\! \frac{3}{8\pi^2(n+1)}\!\bigg[M^2\!-\!(M^2\!+\!s_0)e^{- s_0/M^2}\!\bigg]
\nonumber\\
    &\hspace{0.8cm}  + \frac{\langle\alpha_s G^2\rangle}{8\pi M^2} +\frac{\langle g_s^2\bar qq\rangle^2 (2+\kappa^2)}{486\pi^2M^4}\bigg[35-16\left(-\ln \frac{M^2}{\mu^2}\right)\bigg]
\nonumber\\
    &\hspace{0.8cm}  - \frac{8\,\langle g_s\bar q q\rangle^2}{27M^4} ~+~ \frac{\langle g_s\bar q \sigma TGq\rangle}{M^4}\,+\,3m_q~\frac{\langle\bar q q\rangle}{M^2},
\label{Eq:xi30xi30}
\end{align}
Here, the light quark is taken as $q=(u,d)$ and corresponding vacuum condensates are given in the next Section. Since higher-order and higher-dimensional corrections are difficult to calculate completely, the moments $\langle\xi^n_{2;a_0}\rangle|_\mu$ of the sum rule (\ref{Eq:SRxi2nxi30}) cannot be normalized in the intire Borel parameter region. To get a more accurate moments $\langle\xi^n_{2;a_0}\rangle|_\mu$ for sum rule, we can use the following expression
\begin{equation}
\langle\xi^n_{2;a_0}\rangle|_\mu=\frac{(\langle\xi^n_{2;a_0}\rangle|_\mu\langle\xi^{p;0}_{3;a_0} \rangle|_\mu)|_{\rm From~Eq.~\eqref{Eq:SRxi2nxi30}}}{\sqrt{\langle\xi^{p;0}_{3;a_0}
\rangle^2|_\mu}\big|_{\rm From~Eq.~\eqref{Eq:xi30xi30}}}
\label{Eq:xi2n}
\end{equation}
This method can eliminate the systematic errors caused by many factors. The discussion for the pion and kaon cases can be found in our previous work~\cite{Zhong:2021epq,Zhong:2022ecl}.

Furthermore, $a_0(980)$-meson twist-2 DA $\phi_{2;a_0}(x,\mu)$ describes the momentum fraction distribution of partons in $a_0(980)$-meson for the lowest Fock state. Due to the $\phi_{2;a_0}(x,\mu)$ is the universal nonperturbative physical quantity, it should be researched by the nonperturbative QCD approach. Normally, one can study the $\phi_{2;a_0}(x,\mu)$ based on the combination of nonperturbative QCD and phenomenological model. Meanwhile, conformal expansion of LCDAs Gegenbauer polynomials makes the higher-order Gegenbauer moments unreliable. To improve this situation, the LCHO is adopted to determine $a_0(980)$-meson twist-2 LCDA. Referring to the LCHO model of the pion leading-twist WF raised in Refs.~\cite{Wu:2010zc, Wu:2011gf}, one can start with the Brodsky-Huang-Lepage (BHL) prescription, which assuming there is a connection between the equal-times wave function in the rest frame and the light-cone wave function~\cite{BHL}. it can be expressed as:
\begin{align}
&\Psi_{2;a_0}(x,\textbf{k}_\bot)= \sum_{\lambda_1\lambda_2} \chi_{2;a_0}^{\lambda_1\lambda_2}(x,\textbf{k}_\bot) \Psi^R_{2;a_0}(x,\textbf{k}_\bot),
\label{WF_full}
\end{align}
where $\textbf{k}_\bot$ is transverse momentum. Furthermore, $\lambda_1$ and $\lambda_2$ are the helicities of the two constituent quark. $\chi_{2;a_0} ^{\lambda_1 \lambda_2} (x, \textbf{k}_\bot)$ stands for the spin-space WF that comes from the Wigner-Melosh rotation. The different forms of $\lambda_1\lambda_2$ are shown in Table \ref{table:chi}, which can also been seen in Refs.~\cite{Huang:1994dy, Cao:1997hw, Huang:2004su, Wu:2005kq}. The spin-space WF, e.g. $\sum_{\lambda_1 \lambda_2} \chi_{2;a_0}^{\lambda_1 \lambda_2} (x,\textbf{k}_\bot) ={\hat m_q}^2/(\textbf{k}^2_\bot + {\hat m_q}^2)^{1/2}$. Then, by combing the spatial WF $\Psi^R_{2;a_0}(x,\textbf{k}_\bot)=A_{2;a_0} \varphi_{2;a_0}(x) \exp[ -(\textbf{k}^2_\bot + {\hat m_q}^2)/(8\beta_{2;a_0}^2 x\bar x)]$, the $a_0(980)$-meson WF will be obtained. Here, the $A_{2;a_0}$, $\hat m_q$ stand for the normalization constant and light quark mass. The final $a_0(980)$-meson twist-2 LCDA can be obtained by using the relationship between the twist-2 LCDA and WF of $a_0(980)$-meson, {\it e.g.} integrated over the squared transverse momentum, which can be expressed as
\begin{align}
&\phi_{2;a_0}(x,\mu)=\frac{ A_{2;a_0}\hat m_q \beta_{2;a_0}}{4\sqrt{2}\pi^{3/2}} \sqrt{x\bar x} \varphi_{2;a_0}(x)
\nonumber\\
&\quad\times \left\{ \textrm{Erf}\left[ \sqrt{\frac{{\hat m_q}^2+\mu^2 }{8\beta_{2;a_0}^2 x\bar x}} \right] - \textrm{Erf}\left[ \sqrt{\frac{{\hat m_q^2}}{8\beta_{2;a_0}^2 x\bar x}} \right] \right\},
\label{DA_model}
\end{align}
where ${\rm Erf}(x) = 2\int^x_0 e^{-t^2} dx/{\sqrt{\pi}}$ is the error function, and $\varphi_{2;a_0}(x) = (x\bar x)^{\alpha_{2;a_0}} C_1^{3/2}(2x-1)$. Based on the experience of other mesons~\cite{Huang:2013gra, Huang:2013yya, Wu:2012kw, Zhong:2015nxa, Zhang:2021wnv, Hu:2021lkl, Zhong:2018exo, Zhong:2022ecl, Zhong:2021epq}, we take the wavefunction parameter $\beta_{2;a_0}=0.5$. Whether the value of $\beta_{2;a_0}$ is accurate can be judged by goodness of fit $P_{\chi^2}$. The free parameters $\alpha_{2;a_0}$ and $A_{2;a_0}$ can be obtained by fitting the moments $\langle \xi^n_{2;a_0} \rangle|_\mu$ with the least squares method directly.

\begin{table}
\centering
\renewcommand\arraystretch{1.3}
\small
\caption{The expressions of the spin-space wave function $\chi_{2;a_0}^{\lambda_1 \lambda_2}(x,\textbf{k}_\perp)$ with different $\lambda_1 \lambda_2$.}\label{table:chi}
\begin{tabular}{c c c c}
\\ \hline
$\lambda_1 \lambda_2$ & $\chi_{2;a_0}^{\lambda_1 \lambda_2}(x,\textbf{k}_\perp)$ &
~~~~~~~~~~~~~$\lambda_1 \lambda_2$ & $\chi_{2;a_0}^{\lambda_1 \lambda_2}(x,\textbf{k}_\perp)$\\
\hline
$\downarrow\downarrow$ & $-\dfrac{k_x + i k_y}{\sqrt{2({\hat m_q}^2 + \textbf{k}_\perp^2)}}$ &
~~~~~~~~~~~~~$\uparrow\uparrow$   & $-\dfrac{k_x - i k_y}{\sqrt{2({\hat m_q}^2 + \textbf{k}_\perp^2)}}$ \\
$\uparrow\downarrow$  & $+\dfrac{\hat m_q}{\sqrt{2({\hat m_q}^2 + \textbf{k}_\perp^2)}}$ &
~~~~~~~~~~~~~$\downarrow\uparrow$  & $ - \dfrac{\hat m_q}{\sqrt{2({\hat m_q}^2 + \textbf{k}_\perp^2)}}$ \\
\hline
\end{tabular}
\end{table}

In deriving the complete expression for $D\to a_0(980)$ TFFs, one can take the following correlator
\begin{align}
\Pi_\mu(p,q) = i\int d^4x e^{iq\cdot x}\langle a_0|T \{J_n(x),j_n^\dag(0)\} |0\rangle,
\label{ope}
\end{align}
with $J_n(x)=\bar q_1(x)\gamma_\mu\gamma_5 c(x)$, $j_n^\dag(0)=\bar{c}i\gamma_5q_2(0)$. Followed by the standard LCSR approach, one can make the OPE near the light-cone in the space-like region, and insert a complete set of $D$-meson states in the physical region. After performing the Borel transformation, we can get the $D\to a_0(980)$ TFFs $f_{\pm}^{D\to a_0}(q^2)$ up to twist-3 accuracy, which can be read off
\begin{align}
& f_+^{D\to a_0}(q^2)\!=\!\frac{m_c \bar f_{a_0}}{m_D^2 f_D}\int^1_{u_0} du e^{(m_{a_0}^2- s(u))/ M^2} \bigg\{\!-\!\frac{m_c}u \phi_{2;a_0}(u)
\nonumber\\
&\quad + m_{a_0}\phi_{3,a_0}^p(u)+\frac{m_{a_0}}{6} \bigg[\frac2u \, \phi_{3,a_0}^\sigma(u)-\frac1{m_c^2+u^2m_{a_0}^2-q^2}
\nonumber\\
&\quad \times \bigg((m_c^2 -u^2m_{a_0}^2+q^2) \, \frac{d\phi_{3,a_0}^\sigma(u)}{du} -\frac{4um_c^2m_{a_0}^2}{m_c^2+ u^2m_{a_0}^2-q^2}
\nonumber\\
&\quad \times \phi_{3,a_0}^\sigma(u)\bigg)\bigg]\bigg\},
\label{Eq:Da0-TFF}\\
&f_-^{D\to a_0}(q^2) =\frac{~m_c~ \bar f_{a_0}}{m_D^2 f_D}~\int^1_{u_0} du ~e^{(m_{a_0}^2- s(u))/ M^2} \bigg[\,\frac{\phi_{3;a_0}^p(u)}{u}
\nonumber\\
&\quad +\frac{1}{6u}\frac{d\phi_{3,a_0}^\sigma(u)}{du}\bigg].
\label{Eq:Da0-TFF1}
\end{align}
Here the abbreviation $s(u)=(m_b^2+u\bar{u}m_{a_0}^2-\bar{u}q^2)/u$ is used. The lower limits of the integration is $u_0=\{[(s -q^2 -m_{a_0}^2)^2 +4m_{a_0}^2 (m_c^2-q^2)]^{1/2}-(s-q^2-m_{a_0}^2)\} / (2m_{a_0}^2)$. Here, $m_D$ and $f_D$ are the mass and decay constant of $D$-meson, $m_c$ is the $c$-quark mass, $\bar u = (1-u)$, $s_0$ stands for the continuum threshold. $\phi_{2;a_0}$ and $\phi_{3,a_0}^\sigma$, $\phi_{3;a_0}^p$ are twist-2 and twist-3 LCDAs, respectively. The twist-3 LCDA can be found in~Refs.~\cite{Han:2013zg,Lu:2006fr}. Here, we have a notation that the analytical expressions for the sum rules of the $D \to a_0(980)$ TFFs, {\it i.e}. Eqs.~\eqref{Eq:Da0-TFF} and \eqref{Eq:Da0-TFF1} are equivalent with the tree-level LCSR for the exclusive heavy-to-light $B \to S$ TFFs ($S=a_0(1450), \, K_0^{\ast}(1430), \, f_0(1500)$) have been previously constructed in Ref.~\cite{Wang:2008da}, without the surface term coming from the twist-3 LCDA $\Phi_S^\sigma(u)$ contributions. In this paper, the TFFs analytical expressions are derived based on the pion cases~\cite{Duplancic:2008ix}, where the power of denominator is reduced by taking the derivative of the final state light-cone distribution amplitude.

Then, the explicit expression for the full differential $D\to a_0(980)\ell\bar\nu_\ell$ decay width have the following form
\begin{eqnarray}
\frac{d^2\Gamma (D\to a_0(980)\ell\bar\nu_\ell)}{dq^2 d\cos\theta_\ell} &=& a_{\theta_\ell}(q^2) + b_{\theta_\ell}(q^2) \cos\theta_\ell
\nonumber\\
&+& c_{\theta_\ell}(q^2) \cos^2\theta_\ell,
\end{eqnarray}
with the angular coefficient functions $a_{\theta_\ell}(q^2)$, $b_{\theta_\ell}(q^2)$ and $c_{\theta_\ell}(q^2)$ are \cite{Becirevic:2016hea}
\begin{align}
a_{\theta_\ell}(q^2)&={\cal N_{\rm ew}}\lambda^{3/2}\bigg(1\,-\,\frac{m^2_\ell}{q^2}\bigg)^2\,\bigg[| f_+^{D\to a_0}(q^2)|^2\,+\,\frac1\lambda\frac{m^2_\ell}{q^2}
\nonumber\\
&\times\bigg(1-\frac{m^2_{a_0}}{m^2_D}\bigg)^2|f_0^{D\to a_0}(q^2)|^2\bigg],
\nonumber\\
b_{\theta_\ell}(q^2)&=2{\cal N_{\rm ew}}\lambda\!\bigg(1\!-\!\frac{m^2_\ell}{q^2}\!\bigg)^2\frac{m^2_\ell}{q^2}\!\bigg(\!1-\!\!\frac{m^2_{a_0}}{m^2_D}\bigg){\rm Re}\!\bigg[f_+^{D\to a_0}(q^2)
\nonumber\\
&\times f_{0}^{*D\to a_0}(q^2)\bigg],
\nonumber\\
b_{\theta_\ell}(q^2)&=-{\cal N_{\rm ew}}\lambda^{3/2}\bigg(1-\frac{m^2_\ell}{q^2}\bigg)^3|f_+^{D\to a_0}(q^2)|^2.
\end{align}
Here ${\cal N_{\rm ew}} ={G^2_F|V_{ub}|^2m^3_D} / {256\pi^3}$ with $|V_{cd}|$ and $G_F$ stand for CKM matrix element and fermi coupling constant. $m_\ell$ and $\theta_\ell$ stand for the leptonic mass and helicity angle. In the massless lepton limit, the angular functions $b_{\theta_\ell}(q^2)=0$, $a_{\theta_\ell}(q^2)+c_{\theta_\ell}(q^2)=0$. In addition, $\lambda \equiv \lambda(1,{m^2_{a_0}}/{m^2_D},{q^2}/{m^2_D})$ with $\lambda(a,b,c)\equiv a^2+b^2+c^2-2(ab+ac+bc)$. The $f_{0}^{D\to a_0}(q^2)$ in the expression can be expressed as $f_{0}^{D\to a_0}(q^2)=f_+^{D\to a_0}(q^2)+q^2/{(m^2_B-m^2_D)}f_{-}^{D\to a_0}(q^2)$ \cite{Fu:2013wqa}. After integrating over the helicity angle $\theta_\ell\in[-1,1]$, we can get the expression $q^2$-dependence differential $D\to a_0(980)\ell\bar\nu_\ell$ decay width~\cite{Cheng:2017fkw,Huang:2021owr}
\begin{align}
  &\frac{d\Gamma(D\to a_0(980)\ell\bar\nu_\ell)}{d q^2} = \frac{G_F^2 |V_{cb}|^2}{768 \pi^3 m_D^3} \frac{(q^2\!-\! m_\ell^2)^2}{q^6}\Big[(m_D^2 \!+\! m_{a_0}^2
\nonumber\\
  &~~-q^2)^2-4m_D^2m_{a_0}^2\Big]^{1/2} \bigg\{\bigg[(q^2+m_{a_0}^2-m_D^2)^2(q^2+2m_{\ell}^2)
\nonumber\\
  &~~-q^2m_{a_0}^2(4q^2 \!+\! 2m_{\ell}^2)\bigg](f_+^{D\to a_0}(q^2))^2 \!+\! 6 q^2 m_{\ell}^2 (m_D^2 -m^2_{a_0}
\nonumber\\
  &~~-q^2) f_+^{D\to a_0}(q^2)f_-^{D\to a_0}(q^2)+6q^4m_{\ell}^2 (f_-^{D\to a_0}(q^2))^2 \bigg\}.
\label{wideth}
\nonumber\\
\end{align}

Furthermore, the $D\to a_0(980)$ TFFs are also the basic component of indirect search for new physics Beyond the Standard Model (BSM) phenomenologically. So the angular observables which are sensitive to BSM, {\it i.e.} the normalized forward-backward asymmetries, the $q^2$-differential flat terms and lepton polarization asymmetry ${\cal A}^{D\to a_0(980)\ell\bar\nu_\ell}_{\rm FB}$, ${\cal F}^{D\to a_0(980)\ell\bar\nu_\ell}_{H}$, ${\cal A}^{D\to a_0(980)\ell\bar\nu_\ell}_{\lambda_\ell}$ of the semileptonic decay $D\to a_0(980)\ell\bar\nu_\ell$, respectively. The relationships between these observables and TFFs are as follows~\cite{Cui:2022zwm}:
\begin{align}
{\cal A}^{D\to a_0(980)\ell\bar\nu_\ell}_{\rm FB}(q^2) &= \bigg[\frac12 b_{\theta_\ell}(q^2)\bigg] : \bigg[a_{\theta_\ell}(q^2)+\frac13 c_{\theta_\ell}(q^2)\bigg],
\nonumber\\
{\cal F}^{D\to a_0(980)\ell\bar\nu_\ell}_{\rm H}(q^2) &= \bigg[a_{\theta_\ell} (q^2) + c_{\theta_\ell} (q^2)\bigg] : \bigg[a_{\theta_\ell}(q^2) + \frac13
\nonumber\\
&\times c_{\theta_\ell}(q^2)\bigg],
\nonumber\\
{\cal A}^{D\to a_0(980)\ell\bar\nu_\ell} _{\lambda_\ell}(q^2) &= 1-\frac23\bigg\{\bigg[3a_{\theta_\ell}(q^2) + c_{\theta_\ell} (q^2) + 2m^2_\ell
\nonumber\\
&\times \frac{c_{\theta_\ell}(q^2)}{q^2-m^2_\ell} \bigg]:\bigg[a_{\theta_\ell}(q^2)+\frac13 c_{\theta_\ell}(q^2)\bigg]\bigg\},
\end{align}

\section{Numerical analysis}\label{Sec:III}

To do the numerical analysis, we adopt meson's mass $m_{a_0(980)} = 0.980 \pm 0.020~{\rm GeV}$, $m_D=1.865~{\rm GeV}$ and $m_{D^0}=1.870~{\rm GeV}$. Besides, the current quark-mass are $m_u = 2.16^{+0.49}_{-0.26}~{\rm MeV}$ and $m_d = 4.67^{+0.48}_{-0.17}~{\rm MeV}$ at scale $\mu=2~{\rm GeV}$. The $a_0(980)$-meson decay constant is $f_{a_0}=0.409^{+0.022}_{-0.023}~{\rm GeV}$. The values of the non-perturbative vacuum condensates appearing in the BFTSR are given below as~\cite{Narison:2014ska, Colangelo:2000dp, Zhong:2021epq},
\begin{eqnarray}
\langle \bar qq\rangle &=&( -2.417_{-0.114}^{+0.227} ) \times 10^{-2}~{\rm GeV}^3
\nonumber\\
\langle g_s\bar q\sigma TGq\rangle &=&(-1.934^{+0.188}_{-0.103}) \times 10^{-2}~{\rm GeV}^5
\nonumber\\
\langle g_s\bar qq\rangle ^2&=&(2.082^{+0.734}_{-0.697}) \times 10^{-3} ~{\rm GeV}^6
\nonumber\\
\langle g_s^2\bar qq\rangle ^2&=&(7.420^{+2.614}_{-2.483}) \times 10^{-3} ~{\rm GeV}^6
\nonumber\\
\langle \alpha_s G^2\rangle&=&0.038 \pm 0.011 ~{\rm GeV}^4
\nonumber\\
\langle g_s^3fG^3\rangle&\simeq& 0.045~{\rm GeV}^6
\nonumber\\
\kappa &=& 0.74\pm0.03.
\end{eqnarray}
The values of the double-quark condensate $q\bar{q}$, quark-gluon mixed condensate $\langle g_s\bar q\sigma TGq\rangle$ are
at $\mu = 2~{\rm GeV}$. Otherwise, these parameters can be calculated to any scale according to the evolution equation.

In the context of BFTSR, there are two important parameters the continuous threshold $s_{a_0}$ and the Borel parameter $M^2$, respectively. We generally take the scale $\mu = M$. Under the 3-loop approximate solution, the $\Lambda_{\rm QCD}^{(n_f)} \simeq (324, 286, 207)~{\rm MeV}$ for the number of quark flavors $n_f = 3, 4, 5$, with which the $ \alpha_s(M_z)=0.1179(10)$, $m_c (\bar{m_c}) = 1.27(2)~{\rm GeV}$, $m_b(\bar{m_b})=4.18^{+0.03}_{-0.02}~{\rm GeV}$ and $M_Z = 91.1876(21)~ {\rm GeV}$ are used. Furthermore, the gluon or quark vacuum condensates and non-perturbative matrix element at initial scale can be running to other scales thought out the renormalization group equations RGE)~\cite{Yang:1993bp, Hwang:1994vp}, which can be written as a general formula:
\begin{align}
\chi(\mu) = \bigg[\frac{\alpha_s(\mu_0)}{\alpha_s(\mu)}\bigg]^{y(n_f)} \chi(\mu_0),
\end{align}
with which the function $y(n_f)$ are $-4/b$, $-4/b$ and $-2/(3b)$ for the $m_q$, $\langle \bar qq\rangle$ and $\langle g_s\bar q\sigma TGq\rangle$, respectively. The coefficient $b = (33-2n_f)/3$ and $n_f$ is the number of active quark flavors. According to the basic assumption of BFTSR, it is worth noting that $g_s$ is the coupling constant between background fields in the above vacuum condensates, which is different from the coupling constant in pQCD and should be absorbed into the vacuum condensates as part of these non-perturbative parameters. The RGE of the Gegenbauer moments of the $a_0(980)$-meson twist-2 LCDA is ~\cite{Cheng:2005nb}:
\begin{eqnarray}
a_n^{2;a_0}(\mu) &=& a_n^{2;a_0}(\mu_0) E_n(\mu, \mu_0),
\label{Eq:anRGE}
\end{eqnarray}
with $E_n(\mu, \mu_0) = [\alpha_s(\mu)/\alpha_s(\mu_0)]^{-(\gamma_n^{(0)}+4)/(b)}$. The $\mu_0$ and $\mu$ represent the initial scale and the running scale, the one-loop anomalous dimensions is $\gamma_n^{(0)} = C_F \{[1-2/[(n+1)(n+2)]+4\sum^{n+1}_{j=2}1/j\}$, with $C_F = 4/3$. According to the RGE of the Gegenbauer moments, one can get the moments $\langle\xi^n_{2;a_0}\rangle|_\mu$ for the arbitrary scale $\mu$.
\begin{table}[t]
\centering
\renewcommand\arraystretch{1.3}
\small
\caption{The determined Borel windows and the corresponding $a_0(980)$-meson twist-2 LCDA moments $\langle\xi^n_{2;a_0}\rangle|_\mu$ with $n=(1,3,5,7,9)$ at the scale $\mu_k= 1.4~{\rm GeV}$. Where all input parameters are set to be their central values. In which the abbreviation ``Con.'' indicate the continuum contributions.}
\begin{tabular}{ l c c l}
\\ \hline
$n$~~~~~~& ~~~~~~~~$M^2$~~~~~~~~&~~~~~~~~~~~~$\langle\xi^n_{2;a_0}\rangle|_\mu$~~~~~~~~~~~~&~~~Con. \\
\hline
$1$~ & ~$[2.483,3.483]$~ & ~$[-0.250,-0.203]$~& ~$<30\%$~\\
$3$~ & ~$[1.869,2.869]$~ & ~$[-0.095,-0.129]$~& ~$<20\%$~\\
$5$~ & ~$[3.057,4.057]$~ & ~$[-0.041,-0.074]$~& ~$<40\%$~\\
$7$~ & ~$[4.143,5.143]$~ & ~$[-0.020,-0.048]$~& ~$<60\%$~\\
$9$~ & ~$[4.916,5.916]$~ & ~$[-0.000,-0.027]$~& ~$<75\%$~\\
\hline
\end{tabular}
\label{Tab:tbw}
\end{table}

\begin{figure}[t]
\centering
\includegraphics[width=0.42\textwidth]{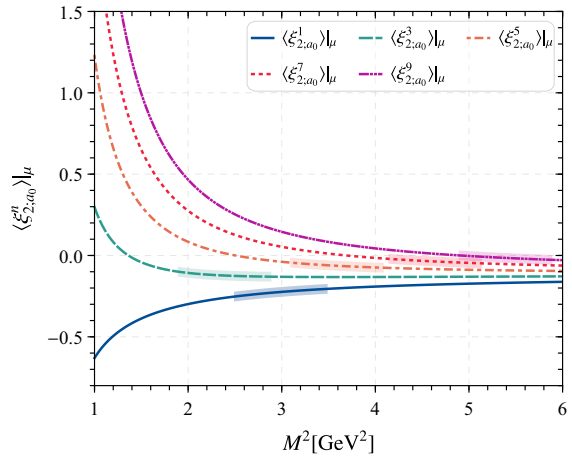}
\caption{The $a_0(980)$-meson leading-twist DA moments $\langle\xi^n_{2;a_0}\rangle|_\mu$ with $n=(1,3,5,7,9)$ versus the Borel parameter $M^2$, where all input parameters are set to be their central values. Where the shaded band indicate the Borel Windows for $n=(1,3,5,7,9)$, respectively.}
\label{Fig:fxinM2}
\end{figure}

Furthermore, the continuum threshold parameter for the sum rule $\langle\xi^n_{2;a_0}\rangle|_\mu$ can be determined by normalization for $\langle\xi^{p;0}_{3;a_0}\rangle|_\mu$, which leads to $s_{a_0}=7~{\rm GeV}^2$. Then, the Borel window for the each order of $a_0(980)$-meson LCDA moments can be determined by limiting the continuum states and the dimension-six condensates contributions. Then, the moments $\langle\xi^n_{2;a_0}\rangle|_\mu$ with $n=(1,3,5,7,9)$ within uncertainties coming from Borel parameters are listed in Table~\ref{Tab:tbw}. Based on the BFTSR, the dimension-six condensates contributions for $\langle\xi^n_{2;a_0}\rangle|_\mu$ are less than $1\%$ for all the $n$th-order. To get the suitable Borel window, the continuum contributions for $\langle\xi^n_{2;a_0}\rangle|_\mu$ are restrict to $(30, 20, 40, 60, 75)\%$ for $n=(1,3,5,7,9)$ respectively. Since the dimension-six condensates contributions are very small, we determine the upper limit of the Borel parameter $M^2$ through the continuum contributions. Then the lower limit of the Borel parameter $M^2$ can be determined by the method of the upper limits, so as to obtain the appropriate Borel window. At the same time, the values of the moments $\langle\xi^n_{2;a_0}\rangle|_\mu$ are stable in the appropriate Borel window. To have a deeper insight into the relationship of the LCDA moments versus Borel parameter $M^2$, the first five moments's curves are shown in Fig.~\ref{Fig:fxinM2}, which can be seen that
\begin{itemize}
 \item In the region $M^2 \in {\rm{[1,2]}}$ of Borel window, the curves for $\langle\xi^n_{2;a_0}\rangle|_\mu$ changed dramatically. With the increase of Borel window $M^2$, the change trend of moments $\langle\xi^1_{2;a_0}\rangle|_\mu$, $\langle\xi^3_{2;a_0}\rangle|_\mu$,
   $\langle\xi^5_{2;a_0}\rangle|_\mu$,
   $\langle\xi^7_{2;a_0}\rangle|_\mu$, $\langle\xi^9_{2;a_0}\rangle|_\mu$ tends to be gentle.
 \item With $n$ increases, the absolute value of the moments $\langle\xi^n_{2;a_0}\rangle|_\mu$ tend to be smaller.
\end{itemize}
\begin{figure*}[t]
\centering
\includegraphics[width=0.42\textwidth]{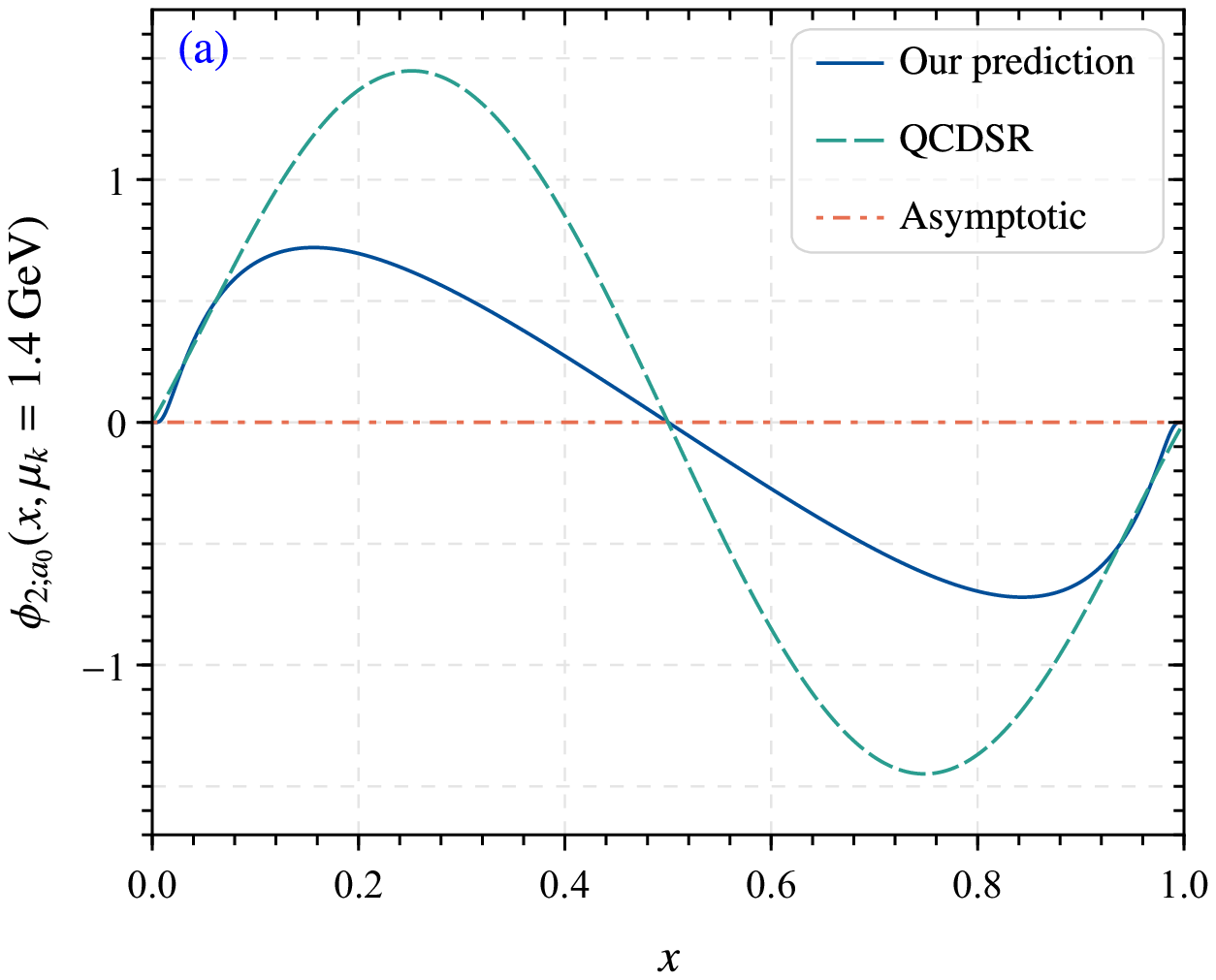}
\includegraphics[width=0.42\textwidth]{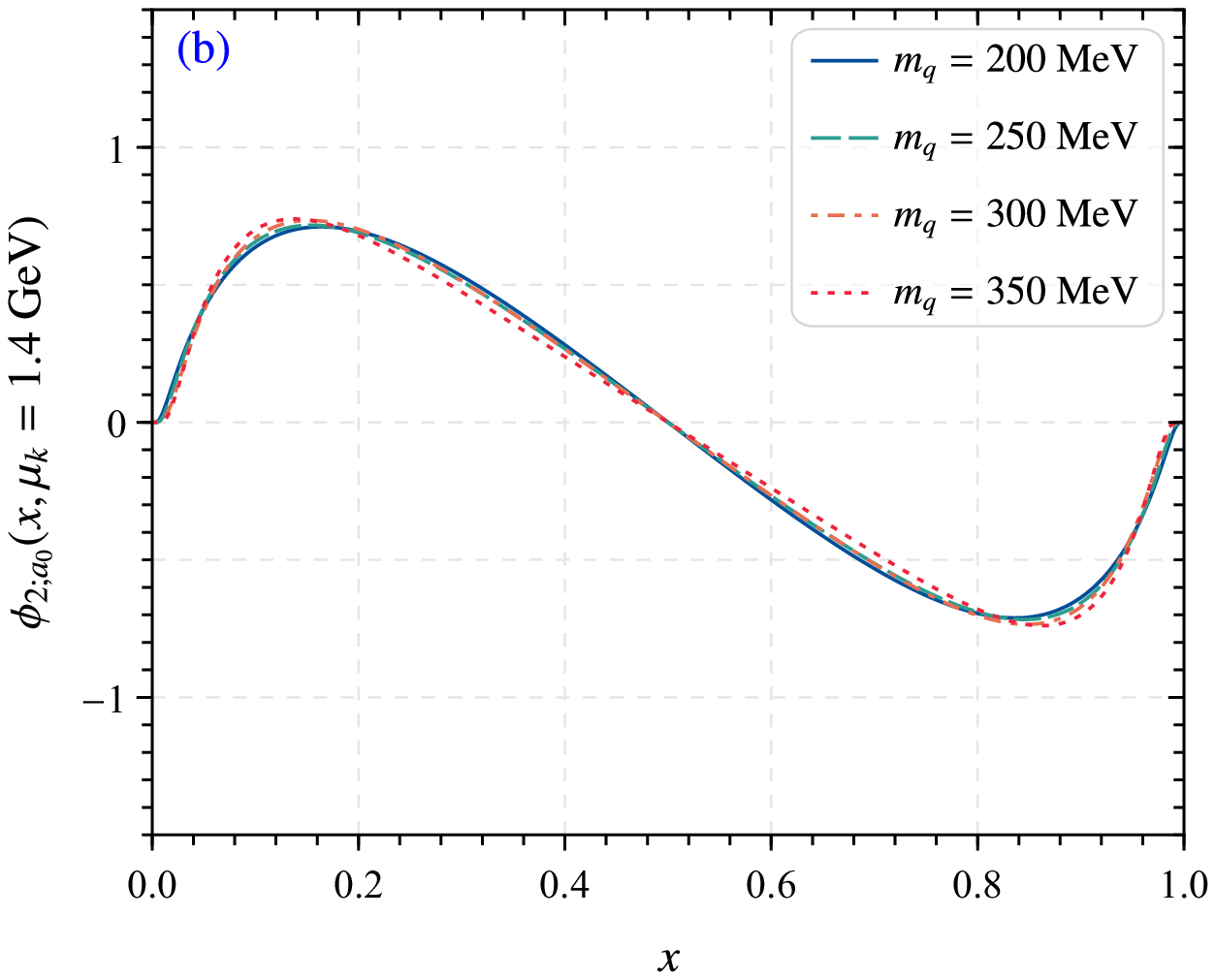}
\caption{The $a_0(980)$-meson twist-2 DA curves in this work. As a comparison, we present the curve from QCDSR prediction~\cite{Cheng:2005nb} in the left panel. In the right panel, the $a_0(980)$-meson twist-2 DA $\phi_{2;a_0}(x,\mu)$ at $\mu_k = 1.4~{\rm GeV}$ with the constituent quark mass $m_q = (200, 250, 300, 350)~{\rm MeV}$ are given, respectively.}
\label{fDAcures}
\end{figure*}
Taking all the input uncertainties into consideration, we can get the moments $\langle\xi^n_{2;a_0}\rangle|_\mu$ with $n=(1,3,5,7,9)$ under two types of factorization scale $\mu_0$ and $\mu_k$,
\begin{align}
&\langle\xi^1_{2;a_0}\rangle|_{\mu_0} = -0.310(43), && \langle\xi^1_{2;a_0}\rangle|_{\mu_k} = -0.250(34), \nonumber\\
&\langle\xi^3_{2;a_0}\rangle|_{\mu_0} = -0.184(32), && \langle\xi^3_{2;a_0}\rangle|_{\mu_k} = -0.126(22), \nonumber\\
&\langle\xi^5_{2;a_0}\rangle|_{\mu_0} = -0.082(27), && \langle\xi^5_{2;a_0}\rangle|_{\mu_k} = -0.067(21), \nonumber\\
&\langle\xi^7_{2;a_0}\rangle|_{\mu_0} = -0.053(25), && \langle\xi^7_{2;a_0}\rangle|_{\mu_k} = -0.044(18), \nonumber\\
&\langle\xi^9_{2;a_0}\rangle|_{\mu_0} = -0.043(23), && \langle\xi^9_{2;a_0}\rangle|_{\mu_k} = -0.024(15),
\label{Eq:xin_mu0}
\end{align}
Our results for the first two order are slightly smaller than the Cheng's predictions, e.g. $\langle\xi^1_{2;a_0}\rangle|_{\mu_0} = -0.56(5)$ and $\langle\xi^3_{2;a_0}\rangle|_{\mu_0} = -0.21(3)$ by using the QCDSR approach from Ref.~\cite{Cheng:2005nb}, which is more likely to be antisymmetric behavior. The little difference may be related to the different methods in determining the continuum threshold $s_{a_0}$. Meanwhile, the higher order such as $n=(5,7,9)$ are given for the first time. The new formulae Eq.~\eqref{Eq:xi2n} can reduce the systematic error of sum rules of the $\xi$-moments, and it enables us to calculate higher-order moments to provide more complete information of DA.
\begin{table}[b]
\centering
\renewcommand\arraystretch{1.3}
\small
\caption{The model parameters $m_q$ (in unit: MeV) of LCHO model $\varphi_{2;a_0}^{\rm IV}(x)$ under different quark masses and their corresponding goodness of fit.}\label{table:fitting}
\begin{tabular}{lllll}
\\ \hline
$m_q$~~~~~~~~~~~~~~~~~~ & $A_{2;a_0}(\rm GeV^{-1})$ & $\alpha_{2;a_0}$~~~~~~~~~~~~~~ & $P_{\chi^2_{\rm min}}$ \\ \hline
200~MeV & $-367$ & $-0.17$ & $0.947$\\
250~MeV & $-239$ & $-0.36$ & $0.954$\\
300~MeV & $-181$ & $-0.51$ & $0.957$\\
350~MeV & $-105$ & $-0.80$ & $0.966$ \\
 \hline
\end{tabular}
\end{table}

\begin{figure}[b]
\centering
\includegraphics[width=0.42\textwidth]{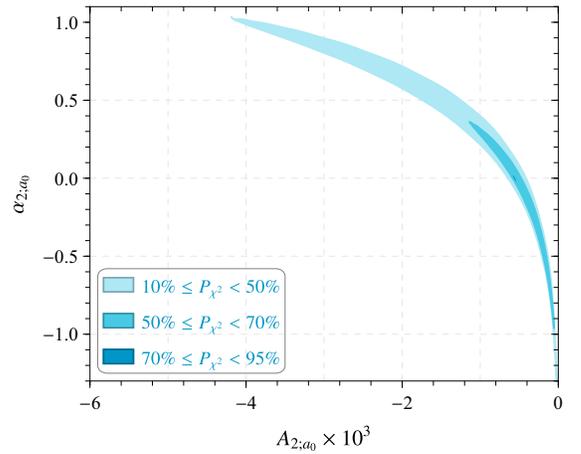}
\caption{Relationship between the goodness of fit $P_{\chi^2_{\rm min}}$ and the two LCHO parameters $A_{2;a_0}$, $\alpha_{2;a_0}$. In which the $P_{\chi^2_{\rm min}}$ is separated into three area with different color.}
\label{Fig:curves}
\end{figure}

\begin{figure*}
\centering
\includegraphics[width=0.42\textwidth]{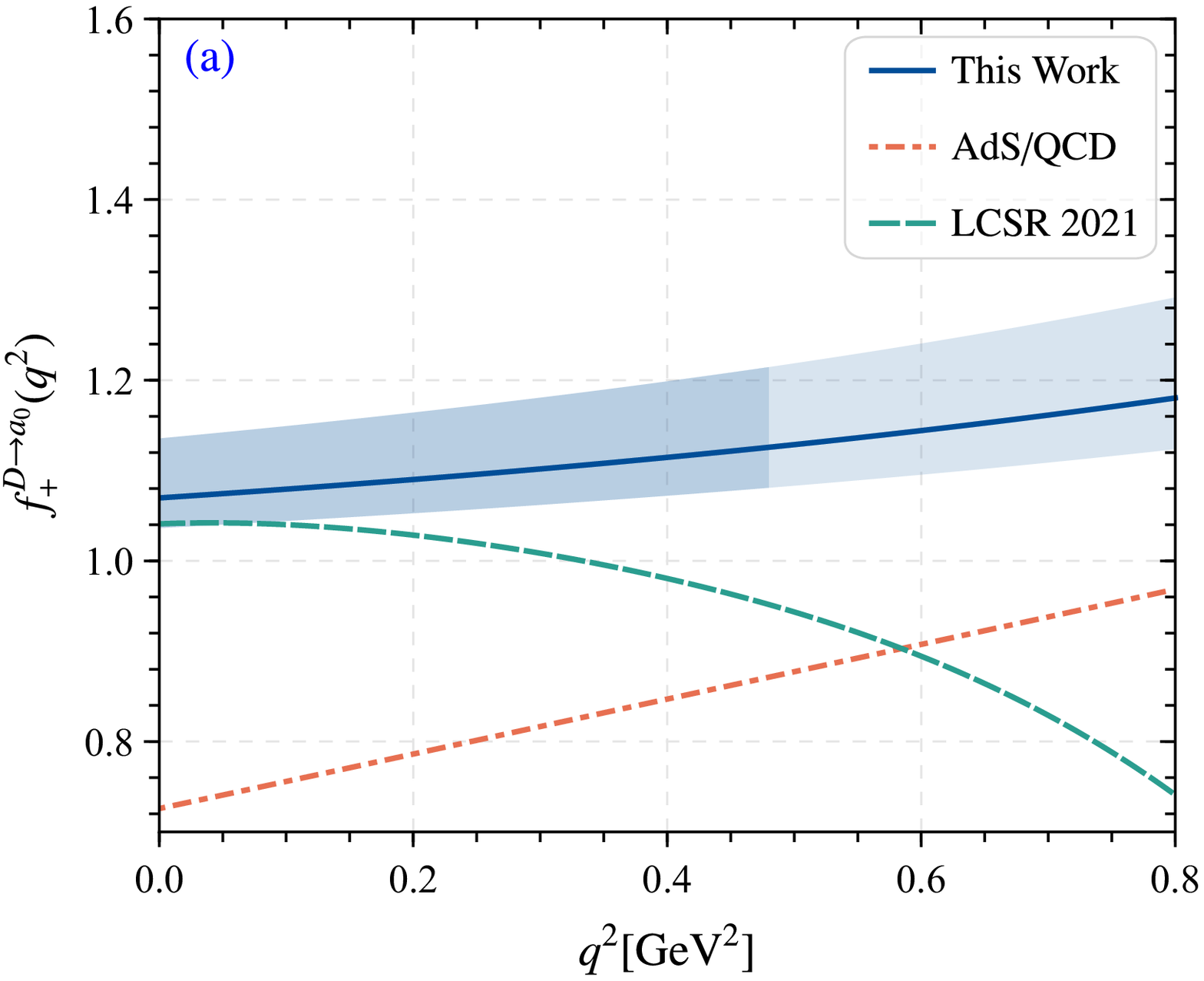}\includegraphics[width=0.42\textwidth]{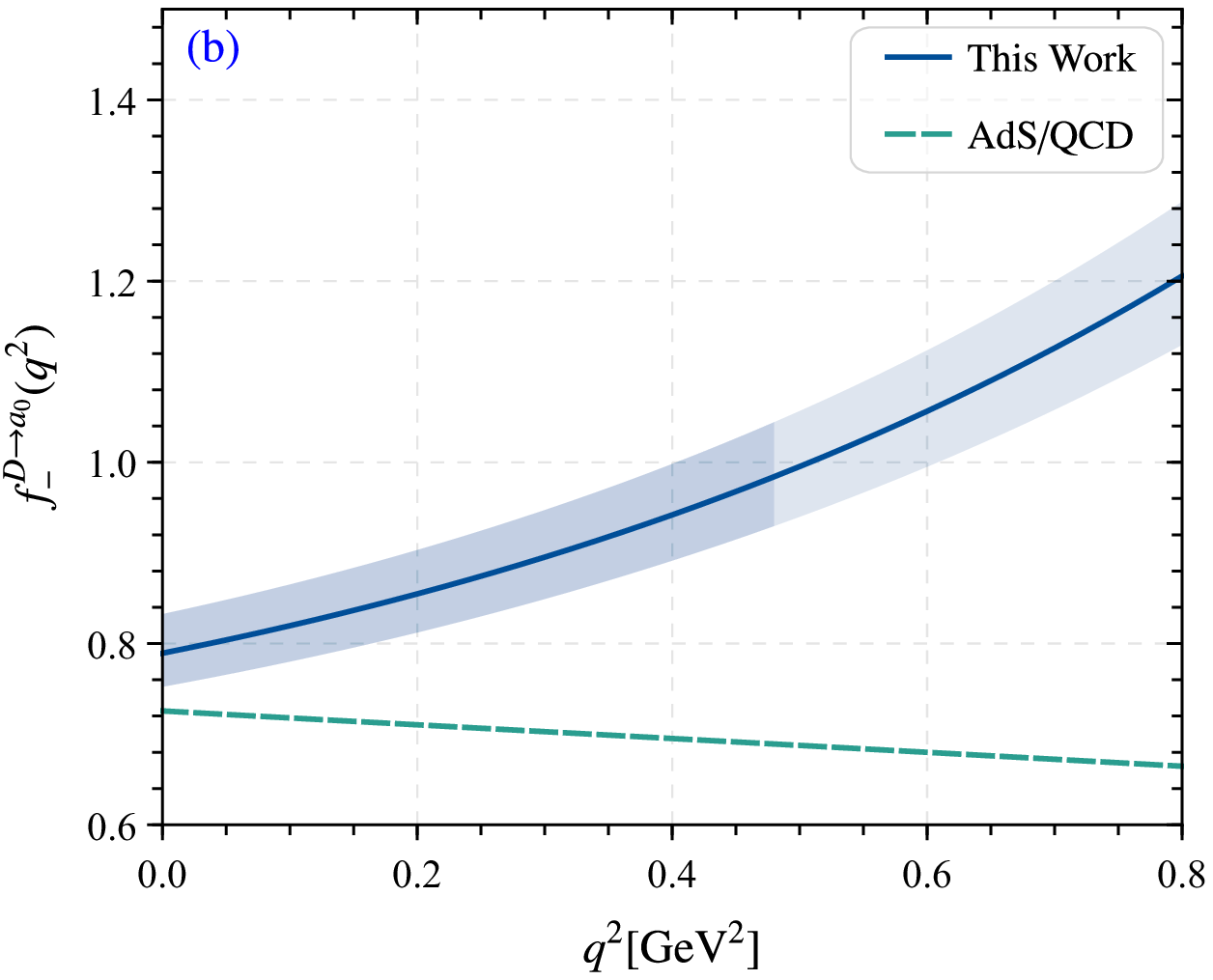}
\caption{The $D\to a_0(980)\ell\bar{\nu}_{\ell}$ TFFs $f_+^{D\to a_0}(q^2)$ and $f_-^{D\to a_0}(q^2)$ of our predictions within uncertainties. Where the darker band stand for the results calculated in LCSR, and the light colored band stand for the SSE. Meanwhile, the results for LCSR~\cite{Huang:2021owr} and Ads/QCD~\cite{Momeni:2022gqb} as a comparison.}
\label{Fig:TFF}
\end{figure*}

Secondly, the free parameters $\alpha_{2;a_0}$ and $A_{2;a_0}$ for the LCHO model can be fixed by adopting the method of least squares to fit $\langle\xi^n_{2;a_0}\rangle|_\mu$ calculated in the framework of the BFTSR shown in Eq.~\eqref{Eq:xin_mu0}. The goodness of fit can be judged by the probability $P_{\chi^2}$ ($P_{\chi^2} \in [0,1]$)~\footnote{For more details, one can see our previous work for pion~\cite{Zhong:2021epq}.}.
By fitting the moments $\langle\xi^n_{2;a_0}\rangle|_{\mu_k}$ at the scale $\mu_k=1.4~{\rm GeV}$, the optimal model parameters are obtained as follows,
\begin{eqnarray}
A_{2;a_0}         &=& -239 ~{\rm GeV}^{-1}, \nonumber\\
\alpha_{2;a_0}  &=& -0.36,         \nonumber\\
\beta_{2;a_0}   &=& 0.5 ~{\rm GeV},
\label{ModelParameterIII}
\end{eqnarray}
The above parameters are derived from $m_q=250~{\rm MeV}$. From these parameters, we obtain the scalar $a_0(980)$-meson twist-2 LCDA $\phi_{2;a_0}(x,\mu_k)$, which is shown in the left panel of Fig.~\ref{fDAcures}. In the meantime, we also present the results of QCDSR~\cite{Cheng:2005nb} prediction as a comparison. From the perspective of variation trend, our prediction results are relatively consistent with the those of QCDSR, and twist-2 LCDA are both antisymmetric. However, there are also some differences between the two, which may be due to differences in calculation methods. The former uses the Gegenbauer moments polynomials expansion, while we take the LCHO model to construct LCDA $\phi_{2;a_0}(x,\mu)$.

For further study, we also analyze the relationship between the goodness-fit of the $a_0 (980)$-meson twist-2 LCDA $\phi_{2;a_0}(x,\mu_k)$ and the quark mass $m_q$ listed in Table~\ref{table:fitting}. It is obvious that as the quark mass increases, the goodness of fit also increases. The goodness of fit $P_{\chi^2_{\rm min}}$ also indicates that $\beta_{2;a_0}=0.5~{\rm GeV}$ is reasonable. Considering the effect of quark mass $m_q$ on the $a_0(980)$-meson twist-2 LCDA $\phi_{2;a_0}(x,\mu_k)$, the behavior of LCHO model $\phi_{2;a_0}(x,\mu)$ for different quark masses $m_q=(200,250,300,350)~{\rm MeV}$ is shown in the right panel of Fig.~\ref{fDAcures}. As can be seen from the figure, the peak value of the LCHO model curves increases with the increase of mass $m_q$. In addition to this, the relationship between goodness of fit $P_{\chi^2_{\rm min}}$ and parameters $A_{2;a_0}$ and $\alpha_{2;a_0}$ is also shown in Fig.~\ref{Fig:curves}. From the table we can see that the $P_{\chi^2_{\rm min}}$ can reach to $95.4\%$, which shows the fitting is good.

\begin{table}[b]
\centering
\renewcommand\arraystretch{1.3}
\caption{TFFs at the point of large recoil $q^2\simeq 0$ for $D\to a_0(980)$ within uncertainties. And a comparison with other theoretical groups are also given.}\label{factor}.
\small
\begin{tabular}{l l l}
\hline
~~~~~~~~~~~~~~~~~~~~~~~~~~~~~~~~&$f_+^{D\to a_0}(0)$~~~~~~~~~~~~~~~~~~& $f_-^{D\to a_0}(0)$ \\
\hline
This~work & $1.070^{+0.066}_{-0.033}$ &	$0.789^{+0.043}_{-0.037}$  \\
CCQM~\cite{Soni:2020sgn} & $0.55^{+0.02}_{-0.02}$ & $0.03^{+0.01}_{-0.01}$ \\
LCSR 2021~\cite{Huang:2021owr} &$0.85^{+0.10}_{-0.11}$	& $\!\!\!\!\!-0.85^{+0.10}_{-0.11}$ \\
LCSR 2017~\cite{Cheng:2017fkw} & $1.76(26)$ & $0.31(13)$ \\
 AdS/QCD~\cite{Momeni:2022gqb}&$0.72(9)$&~~~~-\\
 \hline
\end{tabular}
\end{table}

The TFFs is an important parameter in the calculation of the semileptonic decay $D\to a_0(980)\ell\bar \nu_\ell$. To obtain the numerical results of TFFs, we take $\mu_k = 1.4~{\rm GeV}$, $m_D = 1.869~{\rm GeV}$~\cite{ParticleDataGroup:2022pth}, $f_{a_0(980)} = (0.409^{+0.022}_{-0.023})~{\rm GeV}$~\cite{{Cheng:2005nb}}. For the continuum threshold $s_0$, it is generally taken near the mass square of the first excited state of $D$-meson, that is, near the mass square of $D(2550)$-meson. Based on the prediction of the sum rule of heavy quark effective theory (HQET)~\cite{Huang:1998sa}, we take the continuum threshold parameter $s_0= 6.50(25)~{\rm GeV}^2$. The adopted threshold parameter is consistent with the LCSR analysis for the heavy-to-heavy $B \to D$ form factors with the bottom-meson distribution amplitudes as discussed in Ref.~\cite{Li:2012gr,Wang:2017jow}.

According to LCSR, in the suitable Borel window, we predict the end values of the $D\to a_0(980)\ell\bar{\nu}_{\ell}$ semileptonic decay TFFs shown in Table~\ref{factor}. As a comparison, the results predicted from various approaches, CCQM ~\cite{Soni:2020sgn}, LCSR~2017~\cite{Cheng:2017fkw}, LCSR~2021~\cite{Huang:2021owr} and AdS/QCD~\cite{Momeni:2022gqb} also are presented in Table~\ref{factor}. It is not difficult to see from the Table~\ref{factor} that our predicted results are significantly different from those obtained by other predictions. The reason lies in the twist-2 LCDA $\phi_{2;a_0}(x,\mu)$ is different, which the LCHO model is used here.

\begin{figure*}[t]
\centering
\includegraphics[width=0.42\textwidth]{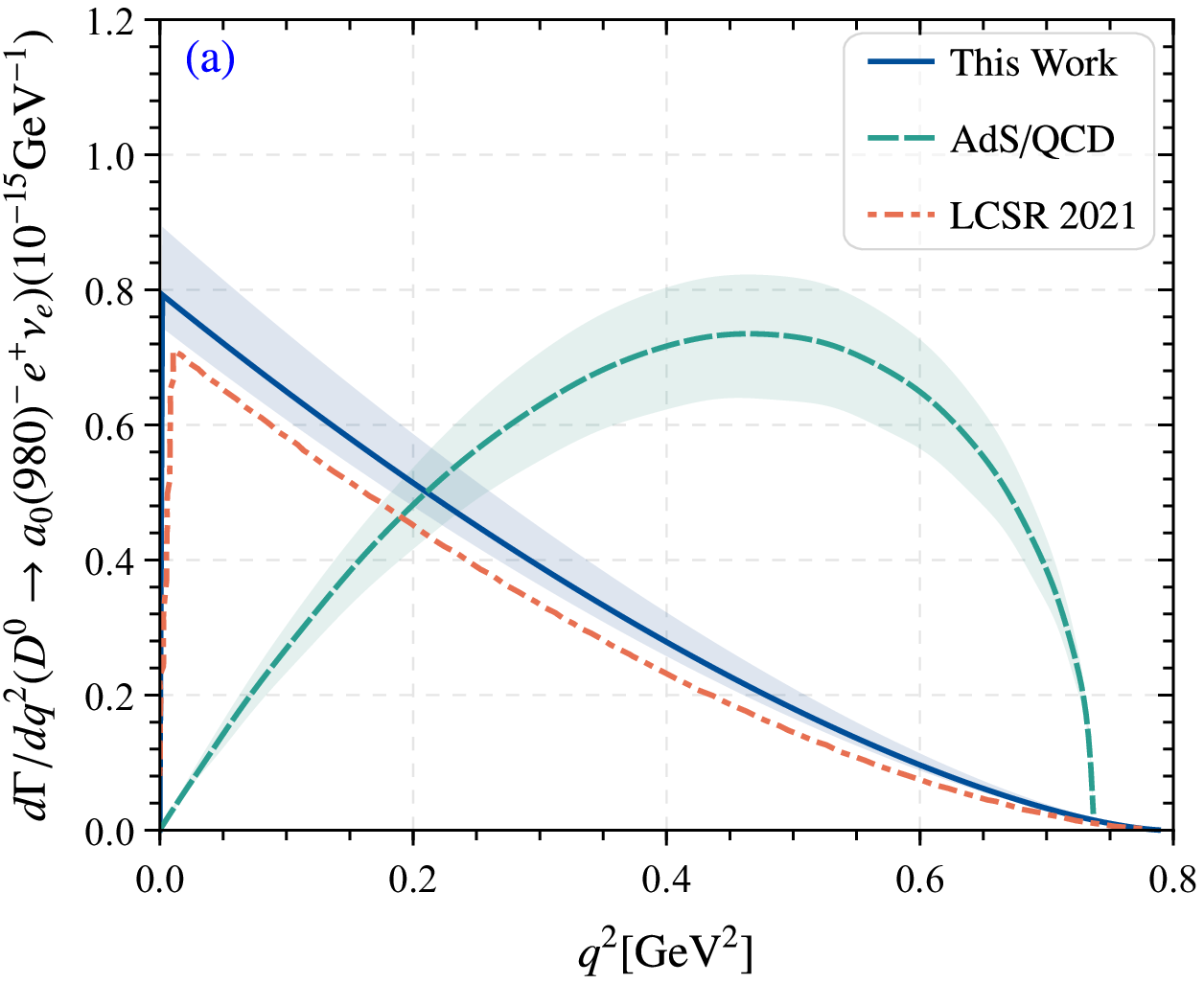}~~\includegraphics[width=0.42\textwidth]{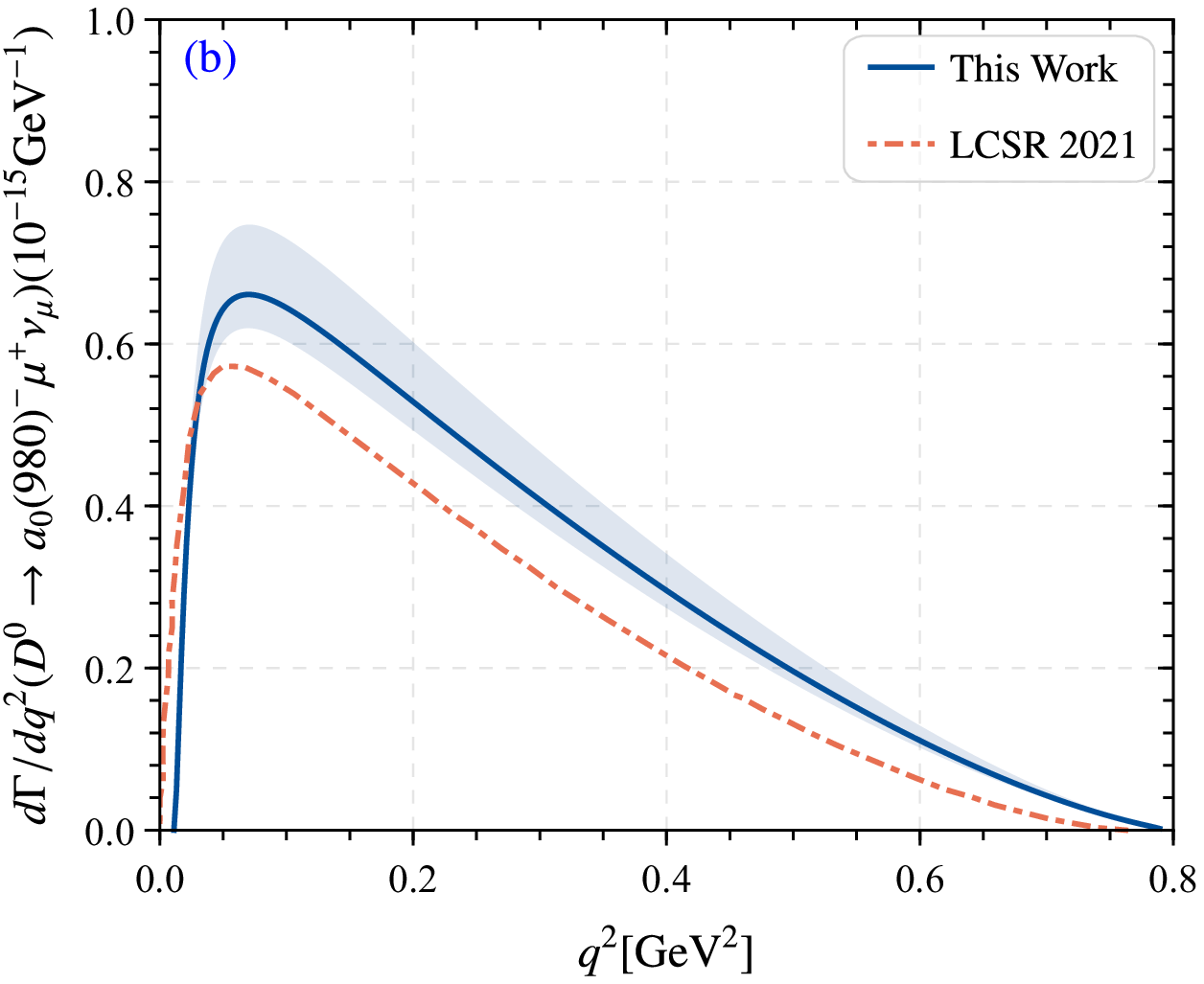}
\includegraphics[width=0.42\textwidth]{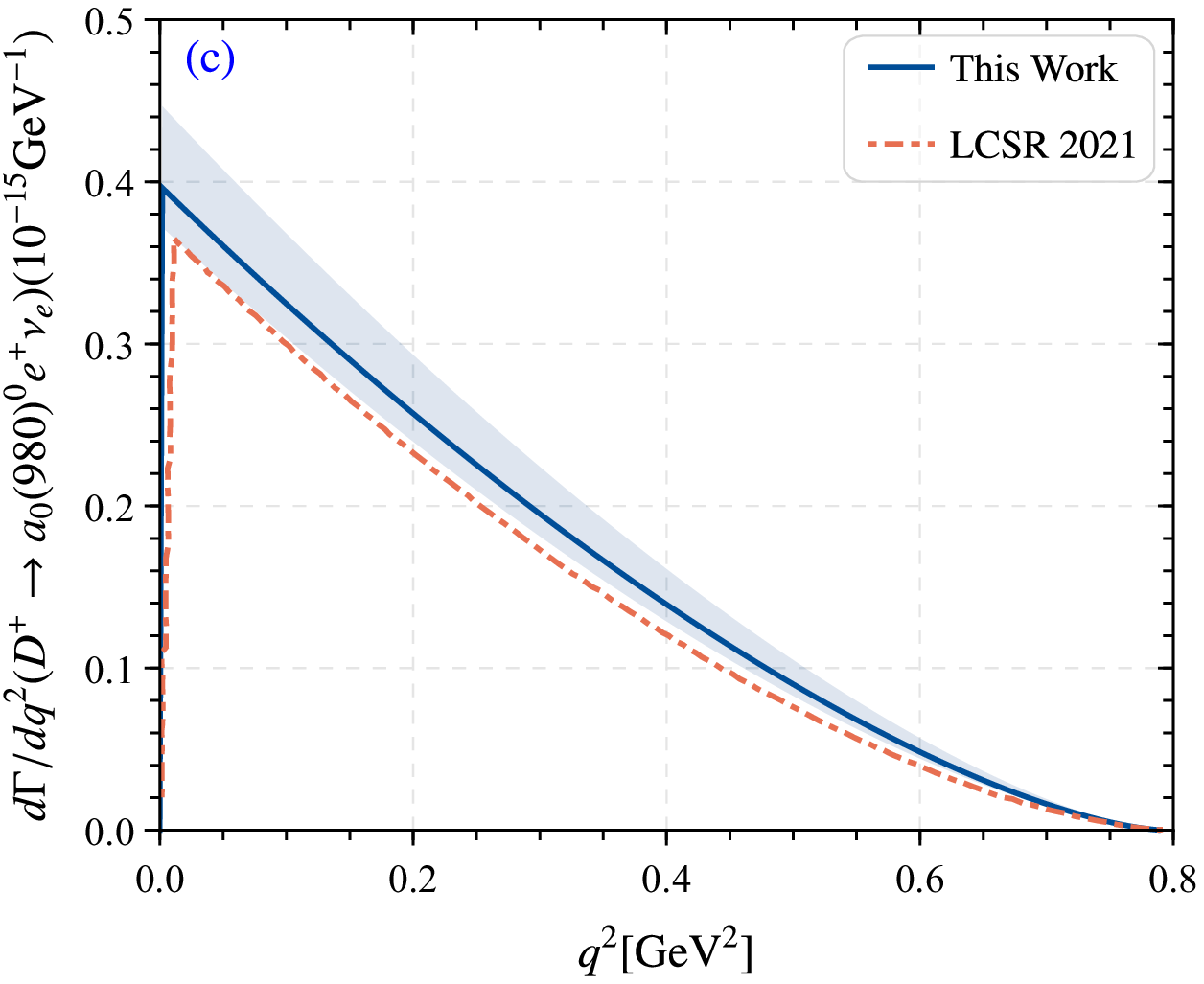}~~\includegraphics[width=0.42\textwidth]{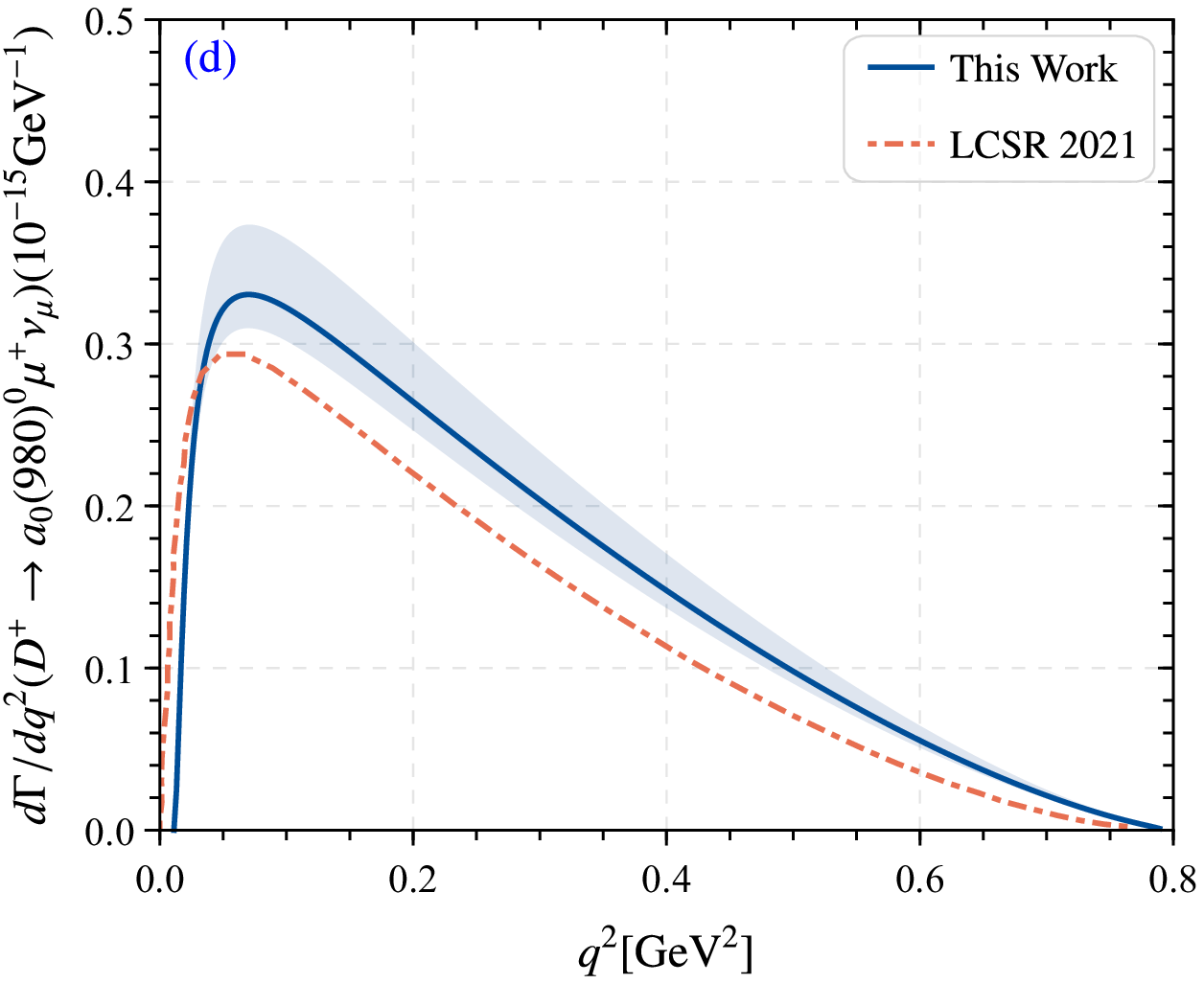}
\caption{Differential decay widths for $D\to a_0(980)\ell\bar{\nu}_{\ell}$ with $\ell = (e,\mu)$ decay process within uncertainties. Where the LCSR~\cite{Huang:2021owr} and AdS/QCD~\cite{Momeni:2022gqb} predictions are also presented.}
\label{width}
\end{figure*}

\begin{table}[t]
\centering
\renewcommand\arraystretch{1.3}
\caption{The fitting parameters $a_i$ with $i=(1,2)$ for TFFs $f^{D\to a_0}_+(q^2)$ and $f^{D\to a_0}_-(q^2)$. Where the goodness of fit $\Delta$ are also present.}\label{Tab:SSE}
\small
\begin{tabular}{l l l l}
\hline
$f^{D\to a_0}_+(q^2)$~~~   & Central value~~~ & Upper limits~~~ & Lower limits \\ \hline
$a_1$     & $6.919$  & $6.312$  & $7.202$  \\
$a_2$     & $29.64$ & $35.97$ & $24.98$  \\
$\Delta$    & 0.18\textperthousand & 0.17\textperthousand & 0.18\textperthousand \\ \hline
$f^{D\to a_0}_-(q^2)$~~~   & Central value~~~ & Upper limits~~~ & Lower limits \\ \hline
$a_1$    & \!\!\!\!\!$-1.793$  & \!\!\!\!\!$-2.069$  & \!\!\!\!\!$-1.396$ \\
$a_2$    & $139.6$ & $158.9$ & $123.0$ \\
$\Delta$  & 0.26\textperthousand & 0.30\textperthousand & 0.23\textperthousand \\
\hline
\end{tabular}
\end{table}

\begin{table*}[t]
\centering
\renewcommand\arraystretch{1.3}
\small
\caption{Branching fractions for the four different channels of $D\to a_0(980)\ell\bar{\nu}_{\ell}$ (in unit: $10^{-4}$). To make a comparison, we also listed the CCQM~\cite{Soni:2020sgn}, LCSR~\cite{Huang:2021owr,Cheng:2017fkw} and AdS/QCD~\cite{Momeni:2022gqb} predictions.}
\begin{tabular}{lllll}
\hline
~~~~~~~~~~~~~~~~~~~~~~~~~~~~~&$ D^0\to {a_0(980)}^{-}e^+\nu_e$~~~~~~~~&$D^0\to {a_0(980)}^{-}{\mu}^+\nu _\mu$~~~~~~~~&$D^+\to {a_0(980)}^0e^+\nu_e$~~~~~~~~&$D^+\to {a_0(980)}^0{\mu}^+\nu_\mu$ \\ \hline
This work    &$1.574^{+0.254}_{-0.156}$ & $1.496^{+0.240}_{-0.147}$& $1.982^{+0.320}_{-0.196}$&$1.885^{+0.302}_{-0.186}$ \\
CCQM ~\cite{Soni:2020sgn} & $1.68\pm{0.15}$  &$1.63\pm{0.14}$  & $2.18\pm{0.38}$ & $2.12\pm{0.37}$ \\
LCSR 2017~\cite{Cheng:2017fkw} &$4.08^{+1.37}_{-1.22}$   & -	& $5.40^{+1.78}_{-1.59}$ &-\\
LCSR 2021~\cite{Huang:2021owr}  &$1.36$   & $1.21$	& $1.79$ &$1.59$\\
AdS/QCD~\cite{Momeni:2022gqb}&$2.44\pm0.30$&-&-&-\\
\hline
\label{branching ratios}
\end{tabular}
\end{table*}
\begin{table*}[t]
\centering
\renewcommand\arraystretch{1.3}
\small
\caption{The absolute branching ratio of $D\to a_0(980)(\to\eta\pi)e^+\nu_e$ (unit: $10^{-4}$) of our predictions within uncertainties. To make a comparison, we also listed the BESIII collaboration~\cite{BESIII:2018sjg}, LCSR results~\cite{Huang:2021owr} and PDG average value~\cite{ParticleDataGroup:2022pth}.}
\begin{tabular}{lllll}
\hline
~~~~~~~~~~~~~~~~~~~~~~~~~~~~~~~&${\cal B}(D^0\to a_0(980)^-(\to\eta \pi^-) e^+\nu_e)$~~~~~~~~~&${\cal B}(D^+ \to a_0(980)^0(\to\eta \pi^0) e^+\nu_e)$ \\ \hline
This work     & $1.330^{+0.216}_{-0.134}$& $1.675^{+0.272}_{-0.169}$ \\
BESIII~\cite{BESIII:2018sjg}        & $1.33^{+0.33}_{-0.29}$ & $1.66^{+0.81}_{-0.66}$ \\
LCSR 2021~\cite{Huang:2021owr}       & $1.15$	& $1.51$ \\
PDG~\cite{ParticleDataGroup:2022pth}        & $1.33^{+0.30}_{-0.29}$ & $1.7^{+0.8}_{-0.7}$ \\
\hline
\label{ratios}
\end{tabular}
\end{table*}

\begin{figure*}[t]
\centering
\includegraphics[width=0.42\textwidth]{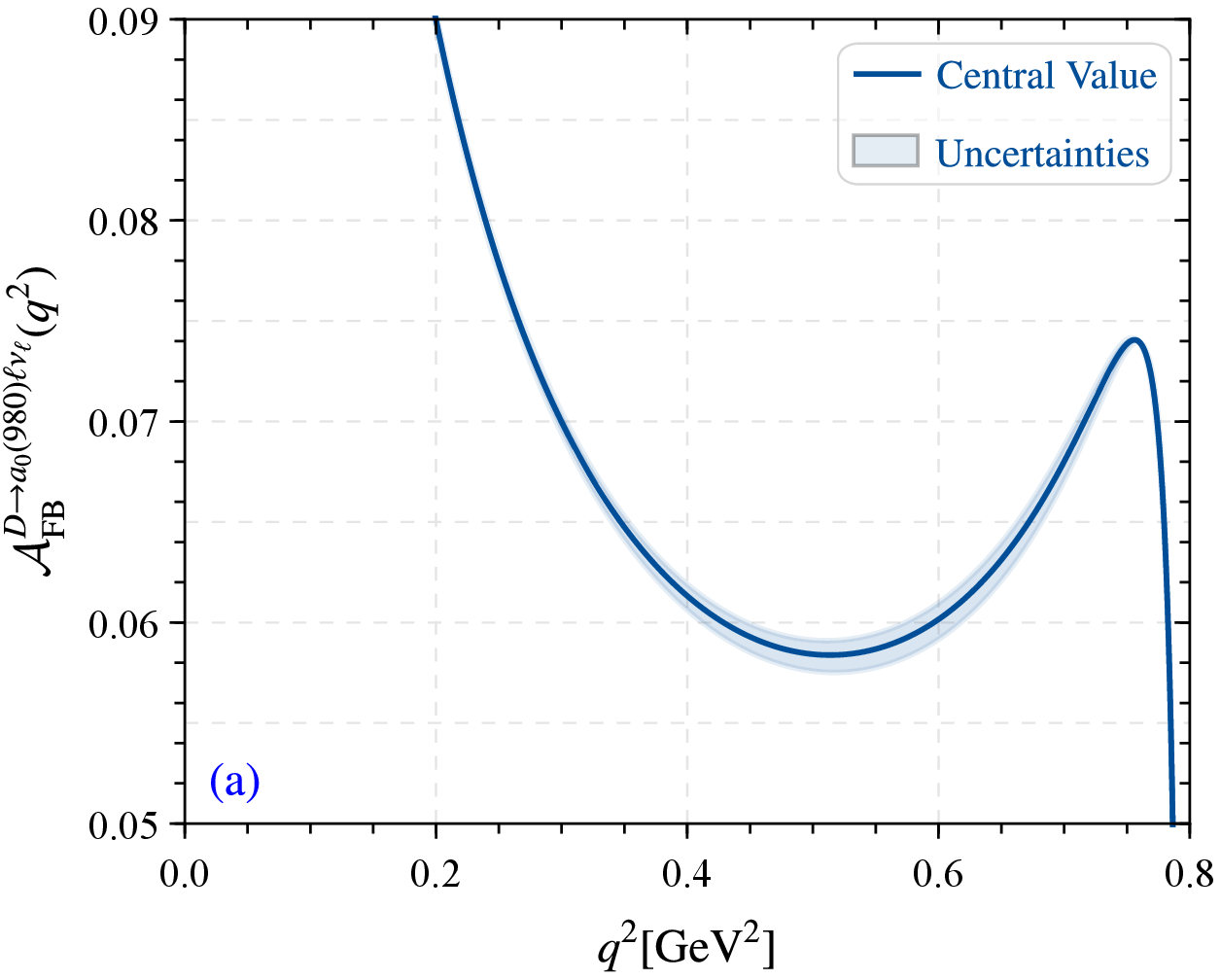}\includegraphics[width=0.42\textwidth]{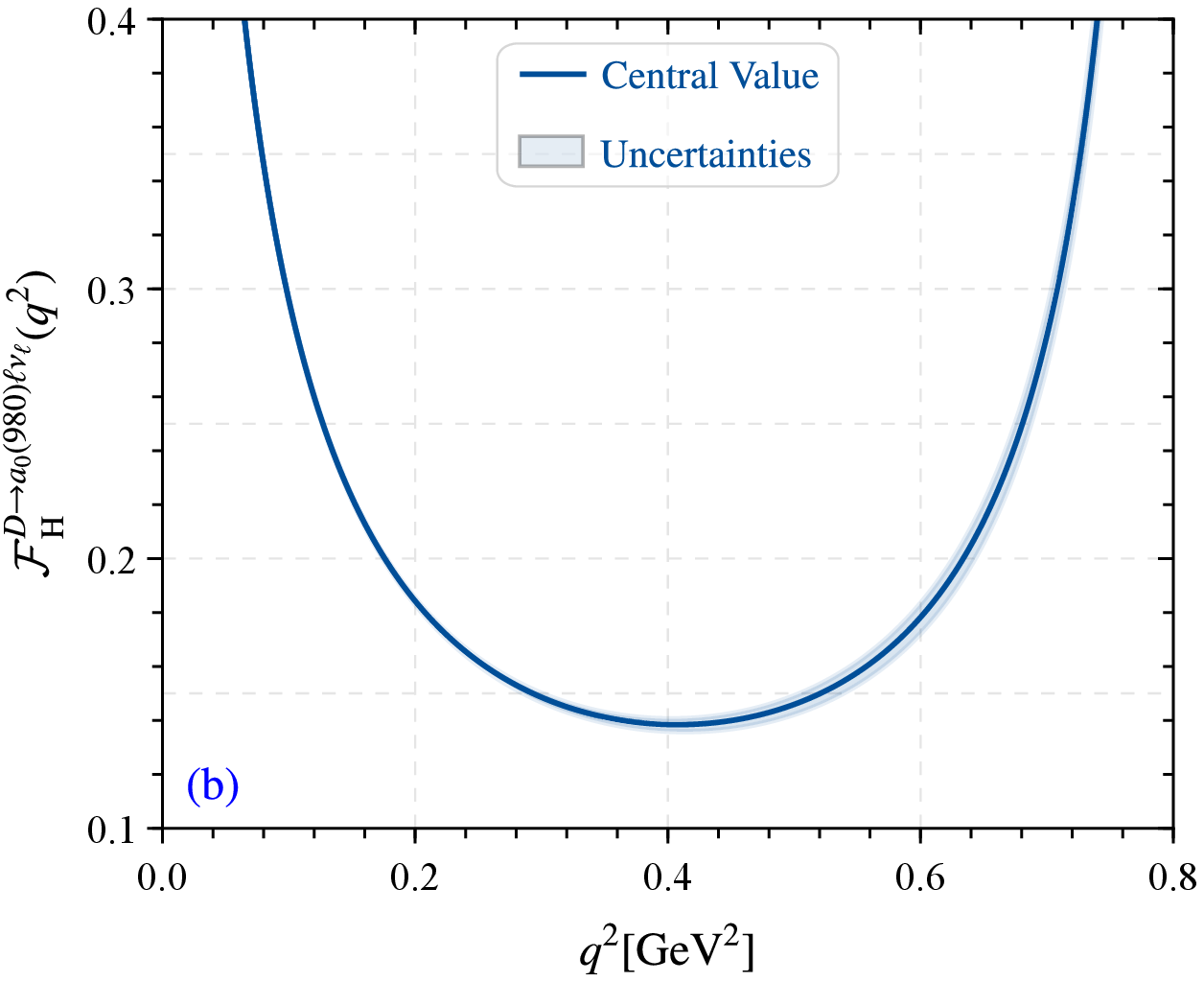}
\includegraphics[width=0.42\textwidth]{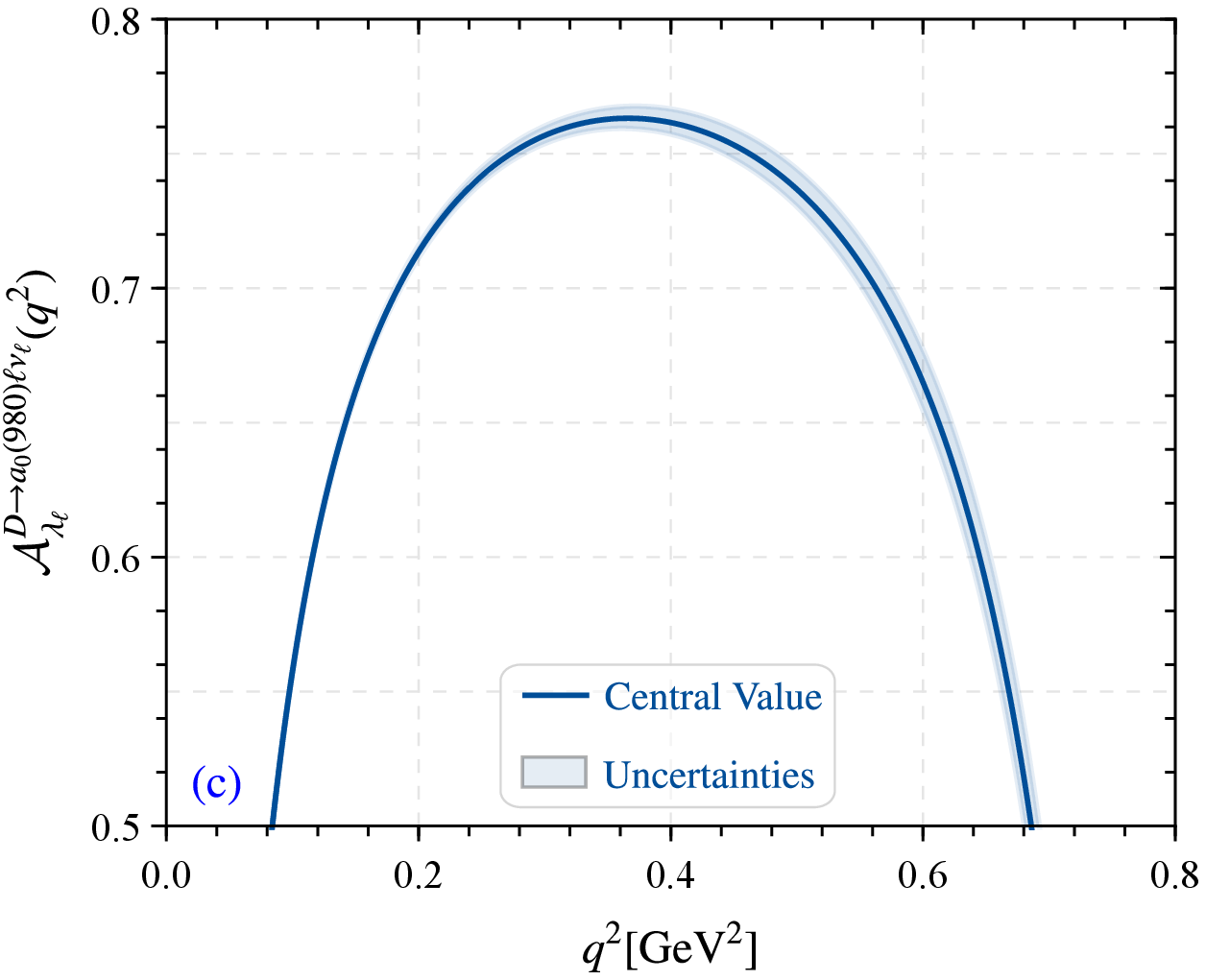}
\caption{The prediction for the three different angular observables ${\cal A}^{D\to a_0(980)\ell\bar\nu_\ell}_{\rm FB}(q^2)$, ${\cal F}^{D\to a_0(980)\ell\bar\nu_\ell}_{\rm H}(q^2)$ and ${\cal A}^{D\to a_0(980)\ell\bar\nu_\ell}_{\lambda_\ell}(q^2)$ within uncertainties, respectively.}
\label{NW}
\end{figure*}

In the physical sense, the LCSR method is suitable for in low and intermediate $q^2$ region. The physical region for $D\to a_0(980)$ TFFs is $m_e\le q^2\le 0.48~{\rm GeV}^2$. In order to obtain reasonable LCSR results, we can extend them to the entire physical $q^2$-region $m_e\le q^2\le (m_D^2-m_{a_0})^2 = 0.8~{\rm GeV}^2$. Therefore, we can fit the complete analysis results with simplified series expansion (SSE), which is a rapidly convergent series on the $z(t)$-expansion~\cite{Fu:2018yin,Bharucha:2015bzk,Bourrely:2008za}.
\begin{align}
{{f}_{i}}(q^2)={P_i}(q^2)\sum\limits_{{k = 0,1,2}} a_k^i {{[z(}{q^2}) - z(0)]}^{k},
\label{ff}
\end{align}
where ${{f}_{i}}(q^2)$ stands for $D\to a_0(980)\ell\bar{\nu}_{\ell}$ TFFs, $a_k^i$ the fit coefficients,and
\begin{align}
P_i(q^2)=\bigg(1-\frac{q^2}{m_{R,i}^2}\bigg)^{-1}
\nonumber\\
z(t)=\frac{\sqrt{t_+-t}-\sqrt{t_+-t_0}}{\sqrt{t_+-t}+\sqrt{t_+-t_0}}
\end{align}
with $t_{\pm}\equiv(m_D\pm m_{a_0})^2$, $t_0=t_+(1-\sqrt{1- t_-/t_+})$. $P_i(q^2)$ is a simple pole corresponding to the first resonance in the spectrum. Meanwhile, there have another alternation version of the $z$-series parametrization for the heavy-to-light TFFs from Wang and Shen~\cite{Wang:2015vgv}, which can also achieve this purpose. The fitting parameters $a_i$ for the TFFs within uncertainties are given in Table~\ref{Tab:SSE}. Meanwhile, the goodness of fit $\Delta = {\sum_t|F_i(t) - F_i^{\rm fit}(t)|}/{\sum_t|F_i(t)|} \times 100$ with $t\in[0,1/100,\cdots,100/100]\times 0.48~{\rm GeV}$ for each TFFs are also presented. After extrapolating TFFs to the whole $q^2$ region, the curves $f_{\pm}^{D\to a_0}(q^2)$ are shown in the Fig.~\ref{Fig:TFF}. We also give the curves obtained by LCSR 2021~\cite{Huang:2021owr} and AdS/QCD~\cite{Momeni:2022gqb} for comparison. The results show our prediction is a certain discrepancy between the results in Ref.~\cite{Huang:2021owr}. There are also some deviations from results predicted by AdS/QCD, but the curve trend is relatively consistent. In most cases, The $f_+^{D\to a_0}(q^2)$ shows an upward trend with the increase of $q^2$. Our prediction is quite reasonable. The $f_-^{D\to a_0}(q^2)$ is also exhibit upward tendency and it turns out that there are some differences.

As other important parameters for $D\to a_0(980)\ell\bar{\nu}_{\ell}$, the CKM matrix element $|V_{cd}|= 0.221\pm 0.008$ from the PDG~\cite{ParticleDataGroup:2022pth}, the fermi coupling constant $G_F=1.166\times10^{-5}~\rm{GeV}^{-2}$. By taking the derived $D\to a_0(980)\ell\bar{\nu}_{\ell}$ TFFs and the related parameters into the differential decay widths Eq.~\eqref{wideth}, one can get the differential decay widths of $D\to a_0(980)\ell\bar{\nu}_{\ell}$ with $\ell= (e,\mu)$ presented in Fig.~\ref{width}. For comparison, we also present the results of LCSR 2021~\cite{Huang:2021owr} predictions in Fig.~\ref{width}. The figure shows that our prediction is consistent with the result of LCSR 2021 within a certain error range. In Fig.~\ref{width}(a), discrepancy between our prediction and the result of AdS/QCD, which may be caused by the different TFFs. Obviously, the predicted results converge to zero in the small recoil region $q^2=(m_D-m_{a_0})^2$. The shaded part in the figure is the error of the width, which mainly comes from the uncertainty of all parameters.

Furthermore, by taking the lifetimes of $D^0$ and $D^+$ meson $\tau_{D^0}=(0.410\pm0.001)~\rm {ps}$ and $\tau _{D^+}=(1.033\pm0.005)~\rm {ps}$ from PDG, we can obtain the branching ratio of semileptonic decay channels $D^+\to {a_0(980)}^0\ell^+\nu_\ell$ ($\ell= e,\mu$) and $D^0\to {a_0(980)}^-\ell^+\nu_\ell$ in Table~\ref{branching ratios}. What's more, the predictions from theoretical groups such as CCQM~\cite{Soni:2020sgn}, LCSR~\cite{Cheng:2017fkw,Huang:2021owr}, AdS/QCD~\cite{Momeni:2022gqb} are shown in Table \ref{branching ratios}. It is obvious that our results are relatively consistent with those of CCQM and LCSR 2021 within the error. However, the significant difference between our predictions and LCSR 2017, AdS/QCD may be caused by the difference of the TFFs and $a_0(980)$-meson twist-2 LCDA.

\begin{table}[b]
\centering
\renewcommand\arraystretch{1.5}
\caption{Numerical results of three observable measurements of semileptonic decay $D\to a_0(980)\mu\bar\nu_\mu$ (the forward-backward asymmetries, the $q^2$ differential flat terms and lepton polarization asymmetry).}\label{observable}.
\small
\begin{tabular}{lll}
\hline
Observables~~~~~~~~~~~~~~~~~~~~~~~~~&Results \\
\hline
${\cal A}^{D\to a_0(980)\mu\bar\nu_\mu}_{\rm FB}$ & $(7.229^{+0.038}_{-0.045})\times10^{-2}$   \\
${\cal F}^{D\to a_0(980)\mu\bar\nu_\mu}_{\rm H}$ &$0.193^{+0.003}_{-0.003}$ \\
${\cal A}^{D\to a_0(980)\mu\bar\nu_\mu}_{\lambda_\ell}$ &$0.460^{+0.007}_{-0.006}$ \\
 \hline
\end{tabular}
\end{table}

Additionally, we also calculate the absolute branching ratio of decays $D\to \eta\pi\ell\bar{\nu}_{\ell}$ by using the relationship
\begin{align}
&{\cal B}(D\to {a_0(980)} (\to \eta\pi)e^+\nu_e) = \nonumber\\
&\quad{\cal B}(D\to {a_0(980)}e^+\nu_e)\times{\cal B}(a_0(980)\to \eta \pi).
\end{align}
Here the ${\cal B}(a_0(980)^0 \to \eta \pi^0)={\cal B}(a_0(980)^- \to \eta \pi^-) = 0.845 \pm0.017$ can be used~\cite{Cheng:2005nb}. Combing with the ${\cal B}(D\to {a_0(980)}e^+\nu_e)$ been calculated in this paper, we can get the results for ${\mathcal B}(D^0\to a_0(980)^-(\to \eta \pi^-) e^+\nu_e)$ and ${\mathcal B}(D^+\to a_0(980)^0(\to \eta \pi^0) e^+\nu_e)$, which are listed in Table~\ref{ratios}. For comparison, the results of the BESIII~\cite{BESIII:2018sjg} collaboration, theory groups LCSR 2021~\cite{Huang:2021owr} and PDG~\cite{ParticleDataGroup:2022pth} predictions are also given. The results show that our predicted absolute branching ratio are in good agreement with those predicted of BESIII, LCSR 2021 and PDG within the error range. It is clear that our prediction is more accurate than the LCSR 2021 prediction, with an improvement of about $12\%$. It shows that the result of our prediction is reasonable.

Finally, the three differential distribution of angle observables for forward-backward asymmetries, the $q^2$-differential flat terms and lepton polarization asymmetry of the semileptonic decay $D\to a_0(980)\ell\bar\nu_\ell$, {\it i.e.} ${\cal A}^{D\to a_0(980)\ell\bar\nu_\ell}_{\rm FB}(q^2)$, ${\cal F}^{D\to a_0(980)\ell\bar\nu_\ell}_{\rm H}(q^2)$ and ${\cal A}^{D\to a_0(980)\ell \bar\nu_\ell}_{\lambda_\ell} (q^2)$ are shown in Fig.~\ref{NW}. The forward-backward asymmetries curves, {\it i.e.} Fig.~\ref{NW}(a) are different with pion and kaon cases~\cite{Cui:2022zwm}, while the $q^2$-differential flat terms and lepton polarization asymmetry have the same tendency overall with pion and kaon cases, but have slight differences. Meanwhile, the uncertainties for the three angle observables are small. Finally, we present the integral results for the three observables in Table~\ref{observable}.

\section{Summary}\label{Sec:IV}

In this paper, we have calculated the moments of $a_0(980)$-meson twist-2 LCDA by adopting the QCDSR approach within the background field theory. The continuum threshold parameter $s_{a_0}$ is determined from the normalization for the $a_0(980)$-meson twist-3 LCDA 0th-order moments. After seeking the suitable Borel windows, we present the first five order moments, i.e. $\langle\xi^n_{2;a_0}\rangle|_\mu$ with $n = (1,3,5,7,9)$ under two different factorization scales $\mu_0$ and $\mu_k$. Then, we study $a_0(980)$-meson twist-2 LCDA $\phi_{2;a_0}(x,\mu)$ based on the LCHO model for improving the accuracy of the calculation. The least square method is used to fit the moments $\langle\xi^n_{2;a_0}\rangle|_\mu$ and to determine the model parameters. The goodness of fit can be up to $95.4\%$. Then, the curves of $a_0(980)$-meson twist-2 LCDA comparing with other theoretical groups and with different constituent quark masses are presented.

The $D\to a_0(980)$ TFFs are calculated within the LCSR approach. The TFFs at large recoil region are listed in Table~\ref{factor}. After extrapolating the TFFs to the whole $q^2$ region, the curves $f_{\pm}^{D\to a_0}(q^2)$ are shown in Fig.~\ref{Fig:TFF}. A comparison of TFFs with other LCSR and AdS/QCD predictions are also given. Using the resultant TFFs, we further studied the semileptonic decays $D\to a_0(980)\ell\bar{\nu}_{\ell}$ with $\ell = (e,\mu)$. Their differential decay widths are presented in Fig.~\ref{width}, and their branching fractions are given in Table~\ref{branching ratios}. The ratio of partial branching fractions is given
\begin{align}
\dfrac{{\cal B}(D^0\to a_0(980)^- e^+ \nu_e)}{{\cal B}(D^+\to a_0(980)^0 e^+ \nu_e)} = 0.794^{+0.155}_{-0.136},
\end{align}
which agree with the CCQM prediction~\cite{Soni:2020sgn} and the LCSR predictions~\cite{Cheng:2017fkw, Huang:2021owr} within errors.

After considering the decay $a_0(980)\to \eta \pi$, we have calculated the branching fractions for the decay processes $D\to \eta\pi e^+\nu_e$, ${\mathcal B}(D^0\to \eta \pi^- e^+ \nu_e) = (1.330^{+0.216}_{-0.134})\times10^{-4}$ and ${\mathcal B}(D^+\to \eta \pi^0 e^+\nu_e)=(1.675^{+0.272}_{-0.169})\times10^{-4}$. The results of our predictions are consistent with the BESIII data and PDG average value within errors, which indicate the two-quark picture of $a_0(980)$ is also reasonable in comparing with four-quark picture. Along with other results of branching fraction for scalar meson discovered experimentally, we will have a reliable input for understanding the nature of the light scalar mesons.

Finally, we also predicted the forward-backward asymmetries, the $q^2$ differential flat terms  and lepton polarization asymmetry of the semileptonic decay $D\to a_0(980)\ell\bar\nu_\ell$. The overall behavior of ${\cal A}^{D\to a_0(980)\ell\bar\nu_\ell}_{\rm FB}(q^2)$, ${\cal F}^{D\to a_0(980)\ell\bar\nu_\ell}_{\rm H}(q^2)$ and ${\cal A}^{D\to a_0(980)\ell\bar\nu_\ell}_{\lambda_\ell}(q^2)$ as a function of $q^2$ is shown in Fig.~\ref{NW}, and the numerical results of the integration are listed in Table~\ref{observable}, which can provides phenomenology value for exploring other new physics.

\acknowledgments
We are grateful for the referee's valuable comments and suggestions. This work was supported in part by the National Natural Science Foundation of China under Grant No.12265010, No.12265009, No.12175025 and No.12147102, the Project of Guizhou Provincial Department of Science and Technology under Grant No.ZK[2021]024 and No.ZK[2023]142, the Project of Guizhou Provincial Department of Education under Grant No.KY[2021]030, and by the Chongqing Graduate Research and Innovation Foundation under Grant No. ydstd1912.

\end{document}